\documentclass[useAMS,usenatbib]{mn2e}
\usepackage{times}

\usepackage{color}
\newcommand{\boldnabla}{\mbox{\boldmath$\nabla$}}

\usepackage{textcomp}
\usepackage{amssymb}
\usepackage{./aas_macros}
\usepackage{graphicx}
\graphicspath{{images/}}

\begin{document}

\title{Magneto-elastic oscillations of neutron stars: exploring different
magnetic field configurations}
\author[Michael Gabler, Pablo Cerd\'a-Dur\'an, Jos\'e
A.~Font, Ewald M\"uller and Nikolaos Stergioulas ]
{Michael Gabler$^{1,2}$, 
Pablo Cerd\'a-Dur\'an$^1$, 
Jos\'e A.~Font$^1$, 
Ewald M\"uller$^2$ 
\and and Nikolaos Stergioulas$^3$ 
\\
  $^1$Departamento de Astronom\'{\i}a y Astrof\'{\i}sica,
  Universidad de Valencia, 46100 Burjassot (Valencia), Spain\\
  $^2$Max-Planck-Institut f\"ur Astrophysik,
  Karl-Schwarzschild-Str.~1, 85741 Garching, Germany \\
  $^3$Department of Physics, Aristotle University of Thessaloniki,
  Thessaloniki 54124, Greece }
\date{\today}
\maketitle
\begin{abstract}
We study magneto-elastic oscillations of highly magnetized
neutron stars (magnetars) which have been proposed as an explanation for the
quasi-periodic oscillations (QPOs) appearing in the decaying tail of the giant 
flares of soft gamma-ray repeaters (SGRs). We extend previous studies by
investigating various magnetic field configurations, computing the 
Alfv\'en spectrum in each case and performing magneto-elastic simulations
for a selected number of models. By identifying the
observed frequencies of $28\,$Hz (SGR 1900+14) and $30\,$Hz (SGR 1806-20) with
the fundamental Alfv\'en QPOs, we estimate the required surface
magnetic field strength.
For the magnetic field configurations investigated (dipole-like poloidal,
 mixed toroidal-poloidal with a dipole-like poloidal component and a 
toroidal field confined to the region of field lines closing inside the star,
and for poloidal fields with an additional quadrupole-like component) 
the estimated dipole spin-down magnetic fields 
are between $8\times 10^{14}$G and $4\times10^{15}$G, in broad 
agreement with spin-down estimates for the SGR sources producing giant
flares.
A number of these models exhibit a rich Alfv\'en continuum revealing 
new turning points which can produce QPOs. This allows one to explain most
of the observed QPO frequencies as associated with magneto-elastic QPOs. 
In particular, we construct a possible configuration
with two turning points in the spectrum which can explain all observed QPOs of
SGR 1900+14. Finally, we find that magnetic field  configurations which
are entirely confined in the crust (if the core is assumed to be a type I
superconductor) are not favoured, due to difficulties in explaining the
lowest observed QPO frequencies ($f\lesssim30\,$Hz).
\end{abstract}
%
\section{Introduction}
The discovery of quasi-periodic oscillations (QPOs) in the decaying tail
of a giant flare of a soft-gamma ray repeater\footnote{SGRs are assumed to
be highly magnetized neutron stars showing repeated bursts in the soft-gamma ray
spectrum \citep{Duncan1992}} (SGR) by \cite{Israel2005} \citep[see also][and
references therein]{Watts2007} has stimulated strong interest in the theoretical
modelling of oscillations of magnetized neutron stars (magnetars). The observed
frequencies of the QPOs are roughly $18$, $26$, $30$, $92$, $150$,
$625$, and $1840\,$Hz for the outburst of SGR 1806-20 and $28$, $53$, $84$, and
$155\,$Hz for SGR 1900+14, respectively.
Since the first attempts to explain these QPOs in terms of torsional,
purely shear modes of the solid crust of the neutron star
\citep{Duncan1998,Strohmayer2005,Piro2005, Sotani2007, Samuelsson2007,
Steiner2009} the theoretical understanding of the
oscillations has evolved. The most recent studies
\citep{Gabler2011letter,Gabler2012,Colaiuda2011, vanHoven2011,vanHoven2012}
favour global magneto-elastic oscillations as an explanation for
the lower frequency ($f<200\,$Hz) QPOs. In these works the coupling through
the magnetic field leads to a very efficient absorption of the shear modes of
the crust into the core even for magnetic field strengths well below those
estimated for
magnetars, $B_{15}\sim 1$ with $B_{15}\equiv B[10^{15} \,\mathrm{G}] $
\citep{Duncan1992}. 
Nonetheless, long-lived QPOs which have predominantly
Alfv\'en character have also been studied
\citep{Sotani2008,Cerda2009,Colaiuda2009}. The frequencies of successive
overtones of these QPOs are found in integer relations and, thus, can
potentially explain some of the observed frequency relations. In two previous 
papers~\citep{Gabler2011letter,Gabler2012} we have shown that these
oscillations can reach the surface of the star with significant amplitudes only
for magnetic fields stronger than $B_{15}\gtrsim 1$, if the internal field
has a global, dipole-like structure.

A successful interpretation of the observed QPOs has the potential to
constrain the equation of state (EoS) of the cold matter at supranuclear
densities which occur only in the core of a neutron star
\citep{Samuelsson2007}. Therefore, neutron
stars are a unique laboratory that can be used to increase our knowledge of
the fundamental physics describing the interaction of nucleons and other elementary
particles at the conditions inside those stars. 

While the current models of magneto-elastic oscillations are promising to explain 
some of the observed QPO frequencies, all existing studies are restricted to a very 
limited set of  magnetic field configurations. The aim of the present work is to 
explore the influence of the magnetic field configuration in the magneto-elastic
model of QPOs to assess its validity. It must be stressed that very little
is known from observations about the internal magnetic field configurations of
magnetars.
The only observational constraint is the presence of a strong external field 
which is responsible for the observed spin-down of the magnetar. Both purely toroidal
and purely poloidal magnetic fields are known to be unstable in unstratified
stars, i.e. stars with a barotropic EoS which
depends on one parameter only \citep{Tayler1973,Markey1973}. Non-linear
simulations confirm these instabilities
\citep[][]{Braithwaite2006, Kiuchi2011, Ciolfi2011, Lasky2011, Lander2011,
Lander2011b}
and suggest that a twisted torus configuration, with mixed poloidal and toroidal field,
is expected. There have been some attempts to model magnetic field configurations
as equilibrium axisymmetric configurations with mixed poloidal and toroidal field
\citep{Colaiuda2008,Kiuchi2008,Ciolfi2009, Lander2009}. However, no stable 
configuration has yet been found for barotropic stars \citep{Lander2012b}.
Nevertheless, different possibilities could stabilize magnetic fields in stars: a more 
complicated magnetic field structure (possibly non-axisymmetric),
stratification, or the presence
of a solid crust. Until this issue is settled, it is legitimate to explore 
the most general setup of magnetic field configurations possible and analyze their 
influence on the QPO frequencies, leaving aside  the stability of the configurations. 
This paper, in which we explore  a wide range of plausible axisymmetric magnetic field 
configurations, summarizes our findings in this direction.

We begin this paper in Section\,\ref{sec_theory} with a short overview of our
theoretical and numerical framework, described in more detail in \cite{Cerda2009}
and \cite{Gabler2012}. We also describe in detail in this section how to obtain different
magnetic field configurations with a new implementation of the {\small MAGSTAR}
tool of the {\small LORENE} library which we call {\small MAGNETSTAR}. In
the following sections we consider
different magnetic field configurations, including dipole-like fields in
Section\,\ref{sec_dipoles} and quadrupole-like fields in
Section\,\ref{sec_quadru}, where  mixed quadrupole-dipole-like
fields are also considered. We further divide the dipole-like fields into purely
poloidal fields
(Section\,\ref{sec_poloidal}) and mixed toroidal-poloidal fields
(Section\,\ref{sec_field_LoreneII}). All magnetic field configurations considered up
to Section\,\ref{sec_quadru} penetrate the whole volume of the star. In 
Section\,\ref{sec_crustfield} we study the case of a magnetic field confined to
the crustal region, a situation motivated by the indications that neutron stars
may contain  superconducting protons \citep{Page2011,Shternin2011} in which case
 the magnetic field could be expelled from the core of the
neutron star, if it is a type I superconductor. A summary of
our results is presented in Section\,\ref{sec_conclusions}.

We use units where $c=G=1$ with $c$ and $G$ being the speed of light
and the gravitational constant, respectively. Latin (Greek) indices
run from 1 to 3 (0 to 3). Partial and covariant derivatives are indicated by a
comma and a semicolon, respectively. We apply the Einstein summation convention.
%
\section{Theoretical framework}\label{sec_theory}
\subsection{General-relativistic magneto-hydrodynamics of elastic bodies}
\label{subsec_mhd}
In the present work we adopt the framework of 3+1 split of general
relativity for a spherically-symmetric spacetime.  The metric takes the
following form
\begin{equation}
 ds^2 = - \alpha^2 dt^2 + \Phi^4 \hat\gamma_{ij} dx^i dx^j \,,
\end{equation}
where $\alpha$ is the lapse function, $\hat\gamma_{ij}$ is the flat,
spatial three-metric, and $\Phi$ is the conformal factor.

The stress-energy tensor $T^{\mu\nu}$ for a magnetized perfect fluid with
elastic properties can be written as a sum of different contributions:
\begin{eqnarray}
 T^{\mu\nu} &=& T^{\mu\nu}_{\,\,\mathrm{fluid}} + T^{\mu\nu}_{\,\,\mathrm{magn}}
+ T^{\mu\nu}_{\,\,\mathrm{elas}}\\
&=&\rho h u^\mu u^\nu + P g^{\mu\nu} + b^2 u^\mu u^\nu + \frac{1}{2} b^2
g^{\mu\nu}- b^\mu b^\nu \nonumber\\
&&- 2 \mu_\mathrm{S} \Sigma^{\mu\nu}\,,
\end{eqnarray}
where $\rho$ is the rest-mass density, $h=1+\epsilon+p/\rho$ the
specific enthalpy, $\epsilon$ the specific internal energy, $P$ the isotropic
fluid pressure, $u^\mu$ the four-velocity of the fluid and $g^{\mu\nu}$ the
metric
tensor. In the magnetic contribution to the stress-energy tensor we
find the magnetic field $b^\mu$, measured by a co-moving observer, with 
$b^2 = b_\mu b^\mu$, and in the elastic contribution we find 
the shear modulus $\mu_\mathrm{S}$ and the shear tensor
$\Sigma^{\mu\nu}$. 

We are interested in torsional oscillations of 
neutron stars and are thus allowed to simplify the model as follows:
(i) We use a {\it barotropic} EoS.
(ii) For {\it poloidal} background fields and in {\it axisymmetry} the {\it axial
oscillations} decouple at the {\it linear level} from the polar oscillations. 
{With these assumptions, 
we evolve only $B^\varphi$ and $S_\varphi$ (see definitions below).
(iii) We apply the {\it Cowling approximation}, regarding the metric  as
fixed.  (iv) For torsional oscillations the shear tensor
$\Sigma^{ij}$ can be expressed as
\begin{equation}
 \Sigma^{ij} = \frac{1}{2}\left[ {\begin{array}{c c c}
0&0& g^{rr} \xi^\varphi_{\,,r}\\
0&0& g^{\theta\theta} \xi^\varphi_{\,,\theta}\\
g^{rr} \xi^\varphi_{\,,r}& g^{\theta\theta} \xi^\varphi_{\,,\theta}   &0\\
\end{array} } \right]\,,
\end{equation}
while the other components vanish, $\Sigma^{\mu 0} = \Sigma^{0 \nu} = 0$. Here,
we have introduced the displacement $\xi^i$ , which is defined via its time
derivative as $\xi^i_{,t}=\alpha v^i$.

With these simplifications the conservation of energy and momentum $\nabla_\nu
T^{\mu\nu}=0$, baryon
number conservation $\nabla_\nu (\rho u^\nu) =0$, and the Maxwell's
equations
$\nabla_\nu \hspace{.5mm} {}^{\ast} \hspace{-1mm} F^{\mu\nu}=0$ and
$\nabla_\nu
F^{\mu\nu}=4\pi \mathcal{J}^\mu$ lead to the following flux-conservative
hyperbolic system of equations
\begin{equation}
 \frac{1}{\sqrt{-g}} \left( \frac{\partial\sqrt{\gamma} {\bf U} }{\partial t} +
\frac{\partial \sqrt{-g} {\bf F}^i}{\partial x^i} \right) = 0\,.
\label{conservationlaw}
\end{equation}
The various quantities introduced here are the Faraday electromagnetic
tensor field $F^{\mu\nu}$ and its dual ${}^{\ast} \hspace{-1mm} F^{\mu\nu}$,
the electric current $\mathcal{J}^\mu$, 
$\sqrt{-g}=\alpha\sqrt{\gamma}$, and $\gamma = det(\gamma_{ij})$. The state
vector $\bf U$ and the flux vector ${\bf F}^i$  are given by
\begin{eqnarray}
 {\bf U} &=& [S_\varphi, B^\varphi]  \label{reduced_withcrust1}\,,\\
 {\bf F}^r &=& \left[ -
\frac{b_\varphi B^r}{W} - 2 \mu_\mathrm{S}
\Sigma^r_{~\varphi}, - v^\varphi B^r
\right]\,,  \label{flux_r}\\
 {\bf F}^\theta &=& \left[ - \frac{b_\varphi B^\theta}{W}- 2 \mu_\mathrm{S}
\Sigma^\theta_{~\varphi},
-v^\varphi B^\theta \right]\,,\label{flux_theta}\label{reduced_withcrust2}
\end{eqnarray}
where $W=\alpha u^t$ is the Lorentz factor, $v^i$ is the three-velocity of the
fluid, and $S_i=(\rho
h + b^2) W^2 v_i - \alpha b_i b^0$ is a generalization of the momentum
density.
The relation between the magnetic field measured by an Eulerian
observer, $B^i$, and $b^i$ is given by
\begin{equation}
 b^\mu = \left[ \frac{W B^i v_i }{\alpha}, \frac{B^i+ W^2 v^j B_j
\hat v^i}{W}\right]\,.
\end{equation}

Eq.\,(\ref{conservationlaw}) is complemented by the evolution equations of the
displacement
\begin{eqnarray} 
(\xi^\varphi_{\,,r})_{,t} - (v^\varphi \alpha)_{,r} &=&0\,,\label{eq_xi_dr}\\
(\xi^\varphi_{\,,\theta})_{,t} - (v^\varphi \alpha)_{,\theta} &=&
0\,.\label{eq_xi_dtheta}
\end{eqnarray}

\subsection*{Boundary conditions}\label{sec_bc}
The boundary and interface conditions have been discussed in
\cite{Gabler2012}. For brevity, here, we only state the conditions and
refer the reader to the previous work for details. The different
conditions are direct consequences of momentum conservation and the
assumption of ideal MHD.

At the neutron star surface this leads to the continuous
traction condition and the assumption that there are no surface currents.
Therefore, the tangential magnetic field components have to be continuous, i.e.
$b^\varphi_\mathrm{crust}=b^\varphi_\mathrm{atmosphere}$ at the
surface and  $\xi^\varphi_{\,\mathrm{crust},r}=0$

At the crust-core interface we have two separate cases:
(i) If the magnetic field penetrates the core we can define a displacement
at this interface as well and $\xi^\varphi$ has to be continuous. As above, we use the
continuity of the traction to set the remaining condition which then reads
$\xi^\varphi_{\,\mathrm{core},r} = \left[ 1+
{\mu_\mathrm{S}}/{\Phi^4 (b^r)^2}
\right]\xi^\varphi_{\,\mathrm{crust},r}$.
(ii) If the magnetic field is confined to the crust, the continuous
traction condition leads to $\xi^\varphi_{\,\mathrm{crust},r} = 0 $ at
the crust-core interface, too.

%
\subsection*{Numerical methods}\label{sec_numerics}
The numerical simulations are performed with the non-linear GRMHD code {\small
MCOCOA} described in previous works
\citep{Cerda2008,Cerda2009,Gabler2012}. This code is able to
compute the torsional oscillations of neutron stars with strong magnetic fields
and an elastic crust. To solve the GRMHD equations we employed a
high-resolution shock-capturing scheme with a method of lines for the
time-advance. 

Complementary to the numerical simulations, we will also
make use of the semi-analytic model presented in
\cite{Cerda2009}, which allows us to compute the spectrum of
standing-wave magneto-elastic
oscillations along individual magnetic field lines. The model is based 
on an integration of a perturbation along the
magnetic field lines in the limit of short wavelengths. We extended this
method in \cite{Gabler2012} to approximately take into account the elastic
crust.

\subsection*{Microphysics}

We construct neutron star models choosing different combinations
of EoS: the EoS APR \citep{Akmal1998} or the stiffer EoS L
\citep{Pandharipande1975} for the core, and  EoS NV \citep{Negele1973} or EoS DH
\citep{Douchin2001} for the crust. 

The calculation of the shear modulus $\mu_\mathrm{S}$ is based on the
zero-temperature limit of \citet{Strohmayer1991}
\begin{eqnarray}
 \mu_\mathrm{S} = 0.1194 \frac{n_i (Ze)^2}{a}\,,\label{shear1}
\end{eqnarray}
where $n_i$ is the ion density, $Ze$ the ion charge, and $a^3 = 3/(4\pi n_i)$
the average ion spacing. For details see \cite{Gabler2012}.

\subsection{Magnetic equilibria}\label{sec_equil_model}

The hydrostatic equilibrium equations for a self-gravitating magnetized star 
\citep[see][and references therein]{Gourgoulhon2012} can be
simplified under the following assumptions: (i) axisymmetry, 
(ii) no rotation or meridional flows, 
(iii) no net charges, (iv) a barotropic EoS of the form $P = P(\rho)$, and
(v) an unstressed crust, $\Sigma^{\mu\nu}=0$. These conditions are appropriate
to describe magnetars, which are regarded to be cold and slowly rotating.
In that case
\begin{equation}
\left ( \log h + \log \alpha \right )_{,i} 
= \frac{1}{\rho h}\epsilon_{ijk} {\mathcal J}^j B^k\, ,
\end{equation}
where ${\mathcal J}^i$ is the magnetic current.
For the magnetic field strength relevant for magnetars, $B < 10^{16}$~G,
deformations of the equilibrium configurations due to  magnetic 
field stresses are small \citep{Bocquet1995}. Therefore, we can expand around the 
unmagnetized spherical equilibrium values, $h=h_0+h'$ and 
$\alpha=\alpha_0+\alpha'$, such that $h'\ll h_0$, $\alpha'\ll \alpha_0$, and 
\begin{equation}
\left (\log h_0 + \log \alpha_0 \right)_{,i} = 0.
\end{equation}
Neglecting magnetic terms in density and pressure terms the resulting
equation for the linearized perturbations is
\begin{equation}
M_{,i} = \frac{1}{\rho_0 h_0}\epsilon_{ijk} {\mathcal J}^j B^k.
\end{equation}
where $M \equiv h'/h_0 + \alpha'/\alpha_0 \ll 1$, which 
quantifies the magnitude of the perturbation and can be used
to assess the quality of the approximation. The equilibrium models
presented later in this work do never exceed a maximum magnitude of
$M\lesssim5\times10^{-4}$. The current and the
magnetic field are related by Ampere's law
\begin{equation}
\epsilon^{ijk}(\alpha B_k)_{,j} = 4 \pi \alpha {\mathcal J}^i \,.
\label{eq:amperelaw}
\end{equation}
Considering the restriction of the solenoidal condition for the magnetic
field there are only two free functions describing the magnetic field
configuration. For convenience we choose them 
to be ${\tilde H}_\varphi \equiv \alpha B_\varphi / 4 \pi$ and $A_\varphi$,  the
latter being the $\varphi$-component of the vector potential
$A_i$ defined as
\begin{equation}
B^i = \epsilon^{ijk}  A_{k,j}.
\end{equation}
The equilibrium equations as a function of $A_\varphi$ and ${\tilde H}_\varphi$ read
\begin{eqnarray}
M_{,r} &=& \frac{1}{\rho_0 h_0} \left[ {\mathcal J}^\varphi A_{\varphi,r} - \frac{4 \pi\gamma^{\varphi\varphi}{\tilde H}_\varphi}{\alpha^2} {\tilde H}_{\varphi,r}\right] \label{eq:equilr} \\
M_{,\theta} &=& \frac{1}{\rho_0 h_0} \left[ {\mathcal J}^\varphi A_{\varphi,\theta} - \frac{4 \pi\gamma^{\varphi\varphi}{\tilde H}_\varphi}{\alpha^2} {\tilde H}_{\varphi, \theta}\right]
\label{eq:equilt} \\
0&=& A_{\varphi,r} {\tilde H}_{\varphi,\theta}  - A_{\varphi,\theta} {\tilde H}_{\varphi,r} \label{eq:equilp}
\end{eqnarray}

\subsubsection{General magnetic field configurations}\label{sec_bfield_general}

Eqs.~(\ref{eq:equilr}-\ref{eq:equilp}) lead to four different branches of
possible
solutions for the current distribution:

\begin{itemize}

\item
 {\it Type I.- Mixed poloidal-toroidal current (${\tilde H}_\varphi = {\tilde H}_\varphi (A_\varphi)$  and $d{\tilde H}_\varphi/dA_\varphi \ne 0$)
} generating a {\it mixed poloidal-toroidal field}: 
Eqs.~(\ref{eq:equilr}-\ref{eq:equilp}) imply $M=M(A_\varphi)$ and 
\begin{eqnarray}
{\mathcal J}^r &=&  \frac{1}{\sqrt{-g}} {\tilde H}_{\varphi, \theta} \nonumber\\
{\mathcal J}^\theta &=&  -\frac{1}{\sqrt{-g}} {\tilde H}_{\varphi, r} \nonumber\\ 
{\mathcal J}^\varphi &=& \rho_0 h_0 \frac{dM}{dA_\varphi} + \frac{4 \pi \gamma^{\varphi\varphi} }{\alpha^2} {\tilde H}_\varphi \frac{d{\tilde H}_\varphi}{d A_\varphi}.
\label{eq:currentmixed}
\end{eqnarray}
In this case we have two free functions ${\tilde H}_\varphi(A_\varphi)$ and
$M(A_\varphi)$ to determine the current ${\mathcal J}^i(A_\varphi)$, which in
turn gives via the $\varphi$-component of
Amp\`ere's law $A_\varphi$. The solution of the system of
equations is identical to that described in \citet{Bocquet1995}. 
It generates a magnetic field with mixed poloidal and toroidal
component. In the case 
$M = 0$, the configuration is force-free.

\item
{\it Type II.-  Purely toroidal current (${\tilde H}_\varphi=0$, $A_\varphi \ne
0$ and ${\mathcal J}_\varphi \ne 0$)}
generating a {\it purely poloidal field}: 
Eqs.~(\ref{eq:equilr}-\ref{eq:equilp}) imply $M=M(A_\varphi)$ and 
\begin{equation}
{\mathcal J}^i = \left ( 0, 0, \rho_0 h_0 \frac{dM}{dA_\varphi} \right )
\end{equation}
For a given form of $M(A_\varphi)$ and using ${\mathcal J}^\varphi$ in the
$\varphi$-component of Ampere's law, Eq.~(\ref{eq:amperelaw}),
an elliptic system results that can be solved for $A_\varphi$. This is
the procedure described by \citet{Bocquet1995} to compute self-consistent
magnetic field equilibria, although their approach is more general since
rotation, charges and the deformations 
of the star are considered, too. This configuration corresponds to a limiting
case of the type I solution for ${\tilde H}_\varphi \to 0$.

\item
{\it Type III.- External currents (${\tilde H}_\varphi = 0$,
${\mathcal J}^\varphi = 0$ and $A_\varphi \ne 0$)}
generating a {\it force-free poloidal field}:
Eqs.~(\ref{eq:equilr}-\ref{eq:equilp}) imply 
$M=0$, i.e. force-free poloidal configurations. It describes
the magnetic field in regions with vanishing currents.
The currents generating the magnetic field are external to the region
described by this currents.
The solution depends only on the potential $A_\varphi$ that is a solution of:
\begin{equation}
\left( \frac{\alpha \gamma_{\theta\theta}}{\sqrt{\gamma}} A_{\varphi,r}\right)_{,r}
+
\left( \frac{\alpha \gamma_{rr}}{\sqrt{\gamma}} A_{\varphi,\theta}\right)_{,\theta} = 0.
\end{equation}
This configuration corresponds to a limiting case of a type II current 
configuration with ${\mathcal J}^\varphi \to 0$.
\item
{\it Type IV.- Purely poloidal current ($A_\varphi=0$ and ${\tilde H}_\varphi \ne 0$)}
generating a {\it purely toroidal field}: 
Eqs.~(\ref{eq:equilr}-\ref{eq:equilp}) imply $M=M({\tilde H}_\varphi)$ and
\begin{equation}
\frac{dM}{d {\tilde H}_\varphi} = -\frac{4 \pi \gamma^{\varphi\varphi} }{\rho_0
h_0 \alpha^2}{\tilde H}_\varphi. \label{eq:toroidal}
\end{equation}
For a given $M({\tilde H}_\varphi)$, Eq.~(\ref{eq:toroidal}) provides the
current and magnetic field configuration
\begin{eqnarray}
{\mathcal J}^i &=& \left ( \frac{1}{\sqrt{-g}} {\tilde H}_{\varphi,
\theta}, -\frac{1}{\sqrt{-g}} {\tilde H}_{\varphi, r} , 0 \right ).
\end{eqnarray}
This current distribution generates a purely toroidal field confined to the
region where ${\mathcal J}^i\ne 0$.

\end{itemize}

Once the current is known, and hence $A_\varphi$ and ${\tilde H}_\varphi$, the
magnetic field can be computed as
\begin{equation}
B^i = \left ( \frac{1}{\sqrt{\gamma}} A_{\varphi, \theta}, -\frac{1}{\sqrt{\gamma}} A_{\varphi,r}, 
\frac{4 \pi \gamma^{\varphi\varphi}}{\alpha}  {\tilde H}_\varphi \right ) \label{eq:potentialfield}
\end{equation}

\subsubsection{Magnetic field in magnetars}\label{sec_bfield_magnetar}

In order to describe magnetars, we make the following assumptions:
(i) the dipole-like component dominates the field at long distances and matches
the observed
spin down within uncertainties. This constraint excludes purely toroidal magnetic fields. 
(ii) There are no currents in the magnetosphere, i.e. the exterior field is
purely poloidal 
and is generated by the currents in the interior (type III). This assumption is
justified by the estimate of \citet{Beloborodov2009}
that any twist in the magnetosphere should be dissipated on a timescale of $\sim
1$~year or shorter.
Under this constraint a toroidal magnetic field can only exist in regions where
magnetic field lines close inside the star (type I), and vanishes otherwise
(type II). We note that some of the work on magnetars
\citep{Colaiuda2008,Lander2012a,Colaiuda2011} use magnetic 
field configurations with a non-zero surface current to mimic
unknown magnetospheric
currents. We discuss below the relevance of these currents and their possible
influence in our simulations.
(iii) There are no current sheets or strong gradients in the magnetar interior
because there is enough evolution time since the magnetar's birth to dissipate
these. Therefore, we favour smooth functions with weak gradients across the
star, i.e. we keep only low-order terms in the free functions
of our magnetic field configurations.

According to these restrictions we consider three magnetic field regions in
magnetars:
(i) the region in the star, where magnetic field lines close in the
interior, can be of type I, II
or III, (ii) the region in the interior where magnetic field lines closing
outside the star, can 
be of type II  only (assuming that the currents generating the field
originate in the interior of the star), and (iii) the exterior region of type
III.
It is convenient to treat all regions within the same framework using the
current given by Eq.~(\ref{eq:currentmixed}) for type I regions, and using the
appropriate limits for ${\tilde H}_\varphi$ and ${\mathcal J}^i$ to construct
the type II and III regions.

We computed the equilibrium models numerically using the 
{\small LORENE} library\footnote{http://www.lorene.obspm.fr}.
To do so we have modified the {\small MAGSTAR} tool of {\small LORENE} to account for
the general form of the current (\ref{eq:currentmixed}) in Ampere's
law~(\ref{eq:amperelaw}). This new routine is called {\small MAGNETSTAR}.
We neglect 
the magnetic field terms in the space-time and hydrostatic equilibrium
equations, and use the same equations and numerical methods as 
in \citet{Bocquet1995}. We first solve the equation for the unmagnetized
equilibrium configuration, and then solve Ampere's law using the spherical
spacetime and the density profile of the unmagnetized star as a background. We
apply a fixed-point iteration to solve Ampere's law, 
starting with an initial current ${\mathcal J}^i_{\rm ini}$. The resulting
magnetic field 
configuration depends on the functions $M(A_\varphi)$ and
$C_\varphi(A_\varphi)$,
and on the initial current $J^i_{\rm ini}$ in the fixed-point iteration.

For $M(A_\varphi)$ we use the form
\begin{eqnarray}\label{eq_poloidal_models}
 M(A_\varphi) &=&  {M_{\rm max}} \times \left[ a_0\, {\tilde
A_\varphi} +
a_1\, {\tilde A_\varphi}^2 + a_2 \,
{\tilde A_\varphi}^3  \right. \nonumber \\
&&\left.+ a_c \, \log\left(1+\frac{{\tilde A_\varphi}}{c}\right)
+ d \, \Theta (A_\varphi - A_\varphi (x)) \right]
\end{eqnarray}
where ${\tilde A}_\varphi \equiv A_\varphi / A_{\varphi}^{\rm max}$
, $A_{\varphi}^{\rm max}$ is the maximum of $|A_\varphi|$ inside the star, 
$A_\varphi (x)$ the value of $A_\varphi$ at the equator at a distance $x$
from the center, $\Theta$ the Heaviside step function, and $a_0$, $a_1$, $a_2$,
$a_c$, and $d$ are constant coefficients, which control the amplitude of the
perturbation $M$, and $c$ a parameter to control the shape of the current.
The field strength of the respective configuration is controlled by $M_{\rm
max}$. 
Note that the expression following $d$
leads to a delta function in the current, and corresponds to a infinitively
thin circular current loop \citep{Jackson1998}. For this
case there exists an analytic solution which we use to avoid dealing with delta
functions numerically.

For ${\tilde H}_\varphi(A_\varphi)$ we use the form
\begin{equation}\label{eq_mixed_models}
{\tilde H}_\varphi(A_\varphi) = b_0 {\hat A}_{\varphi} + b_1 {\hat
A}_{\varphi}^2 + b_2 {\hat A}_{\varphi}^3\,,
\end{equation}
where ${\hat A}_{\varphi} = (A_{\varphi} - A_{\varphi}^{\rm surf. max.}) / 
A_{\varphi}^{\rm max.}$ with $A_{\varphi}^{\rm surf. max.}$ being the maximum
value 
of $A_{\varphi}$ at the surface. The resulting $B_\varphi$ is zero at the
surface, and thus it is confined inside the star. The case $n=1$ corresponds to
one of the configurations in \cite{Lander2012b}.

Regarding the initial current $J^\varphi_{\rm ini}$ we consider three
configurations
\begin{itemize}
\item[] spherical: ${\mathcal J}^\varphi_{\rm ini} = \rho h j_0\,,$
\item[] dipole-like: ${\mathcal J}^\varphi_{\rm ini} = \rho h j_0 r \sin
\theta\,,$
\item[] quadrupole-like: ${\mathcal J}^\varphi_{\rm ini} = \rho h j_0 r^2 \sin
\theta \cos \theta\,,$
\end{itemize}
where $j_0$ is a constant parameter of the same order of magnitude as the coefficients used
for the current ($a_0$, $a_1$ ...). The resulting magnetic field does
not depend on the value of $j_0$ itself, and its strength is determined by the
coefficients of Eqs.\,(\ref{eq_poloidal_models}) and (\ref{eq_mixed_models}),
i.e. $a_i$, $b_i$, $a_c$, and $d$, respectively.
 
Both the spherical and the dipole-like initial currents have even symmetry
with respect to the equator. This also holds when they are rescaled to the
physically relevant $J_{(\varphi)}=\sqrt{ {\mathcal
J}_\varphi {\mathcal
J}^\varphi}$. This $J_{(\varphi)}$ has a maximum at the
equator and resembles a current loop. The corresponding magnetic
field configurations are similar to that of a simple current loop (compare
with the models in Sec.\,\ref{sec_ring_current}). Since the current loop is an
idealization giving rise to a pure dipole field in the far field, we
characterize the
magnetic field configurations generated by the spherical or dipole-like currents
as dipole-like. We note that in the near field regime, where the currents
have finite spatial extension, other multipoles are always present.
Both the spherical and the dipole-like initial current lead to the same final 
equilibrium configurations.
Similarly, we call the magnetic field of a quadrupole-like initial current,
which has odd symmetry with respect to the equator, a quadrupole-like magnetic
field. In this case the dipole moment vanishes and the far field is dominated by
the quadrupole component. To obtain these quadrupole-like configurations one
must use those currents that are an odd function of $A_\varphi$.
Otherwise the fixed-point iteration converges to the same dipole-like
magnetic field as in the previous case. We denote by D and Q the configurations
where the converged magnetic field is predominantly dipole-like or
quadrupole-like, respectively.

Note that for the magnetic field configurations with non-vanishing parameters
$a_0$, $a_1$, and $b_0$ terms appear in the current
that are linear or constant in $A_\varphi$. Therefore, solutions corresponding to different 
parameters $a_0$, $a_1$, and $b_0$, can be linearly superimposed in the current
to obtain further solutions. 
Using this procedure it is possible to compute configurations with a mixed
dipole-quadrupole-like
magnetic field (see Section~\ref{sec_quad}).

We computed most of our initial models using the {\small LORENE}
library and the
procedure described above.
However, in some cases where analytical solutions are known, we use 
these directly: the magnetic field originating from a ring current at
a prescribed radius (see Section\,\ref{sec_ring_current}), and force-free configurations 
confined to the crust according to \cite{Aguilera2008} (see Section\,\ref{sec_crustfield}).

%
\section{ Dipole-like configurations}\label{sec_dipoles}
\begin{figure*}
\begin{center}	
\includegraphics[width=.99\textwidth]{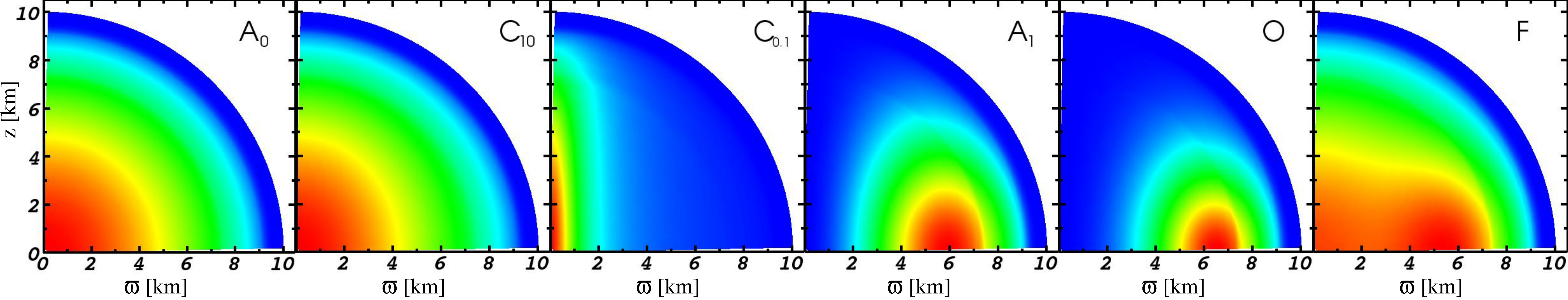}
\includegraphics[width=.99\textwidth]{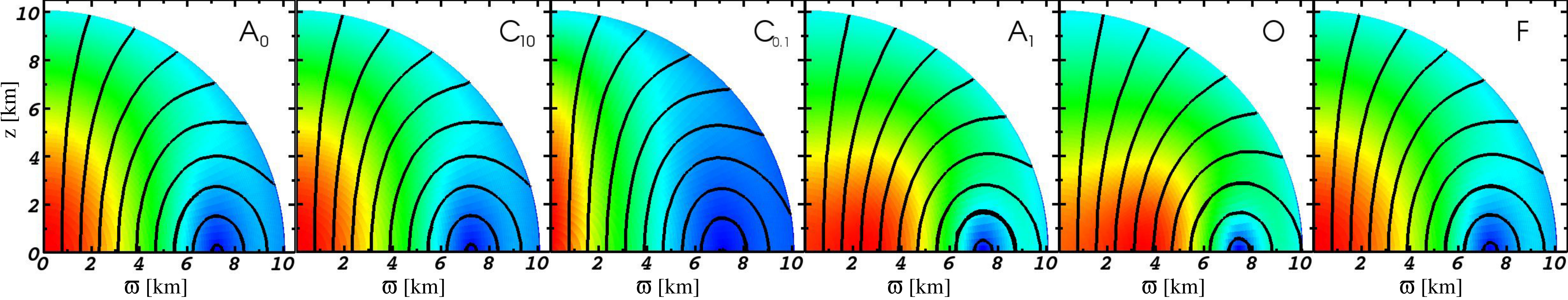}
\end{center}
\caption{{\it Upper panels:} Current distribution $\mathcal{J}^\varphi$ for the
model configurations
(from left to right) A$_0$, C$_{10}$,C$_{0.1}$, A$_1$, O, and F . {\it Lower
panels:} Corresponding magnetic field configurations with selected field lines. The colour
scale ranges from blue (minimum) to red (maximum) and represents the current
strength (top panels) and the strength of the poloidal magnetic field (bottom
panels), respectively. Here and henceforth we use cylindrical coordinates
$z$ and $\varpi$ to indicate the height and radial distance from the $z$-axis,
respectively.}
\label{fig_dipoles}
\end{figure*}

In this section we study the effects of different dipole-like
magnetic field configurations on the Alfv\'en spectrum of a magnetar. We
consider either a purely poloidal component (Section~\ref{sec_poloidal}) or a
mixed poloidal-toroidal field whose poloidal
component is dipolar-like (Section~\ref{sec_field_LoreneII}).
Our analysis is mainly based on the application of the semi-analytic model
to investigate the behaviour of purely toroidal oscillations. Using this
model we compute the Alfv\'en frequencies of individual field lines. The
spectrum of the star is determined by the set of the frequencies of all field
lines and the corresponding overtones given as a function of position
inside the star. The position can be described by the point  $\chi$ at
which a field line crosses the equatorial plane or by the polar angle
$\theta$ along a given radius.
We also employ some numerical simulations to evaluate the quality of our
approach.

\begin{table}
\begin{center}
\begin{tabular}{c | c c c c c | |c}
Model & $a_0$ & $a_1$ &$a_2$ & $a_c$ & $c$ & Radius of \\
&&&&&&ring in km\\ \hline
A$_0$ & 1 &0&0&0&-&-\\
C$_{c}$ &0&0&0&1&$c$&-\\
A$_1$ & 0 &1 &0&0&-&-\\
O&0&0.43&0.57&0&-&-\\
F&0.79&0&0.21&0&-&-\\ \hline
R$_\varpi$&-&-&-&-&-&$\varpi$
\end{tabular}
\caption{Values of the current parameters of the employed models
according to Eq.\,(\ref{eq_poloidal_models}). The models R$_\varpi$ represent
ring currents, $\varpi$ denoting the radius of the ring in km.
}\label{tab_dipole_models}
\end{center}
\end{table}

\subsection{Purely poloidal fields (type II)}\label{sec_poloidal}

We first discuss magnetar configurations with a purely toroidal current (type
II).
We generate the initial models using the procedure described in
Sections\,\ref{sec_bfield_general} and \ref{sec_bfield_magnetar}, and the
{\small LORENE} library. We focus on type II magnetic fields, which
are completely defined by Eq.\,(\ref{eq_poloidal_models}). The corresponding
model parameters are given in Table\,\ref{tab_dipole_models}. We obtain the
desired magnetic field strength by rescaling $M_{\rm max}$.
In the first three models (A$_0$, A$_1$, and C$_{c}$) only one parameter in
Eq.\,(\ref{eq_poloidal_models}) is different
from zero ($a_0$, $a_1$, or $a_c$ respectively). Additionally, we consider two
different combinations of these parameters which are physically motivated.
One of these, model F, results in a very flat spectrum and is
expected to produce long-lasting QPOs and large gaps between the spectra of
adjacent overtones of the Alfv\'en oscillations. Another model, model O, is
chosen such that its spectrum
reproduces some of the observed frequency ratios of SGR 1900+14.

The currents $\mathcal{J}^\varphi$ and the magnetic field configurations of the
different models are
displayed in Fig.\,\ref{fig_dipoles} in the top and bottom panels, respectively.
The different currents lead to very similar magnetic field
configurations, all having their maximum either at the center of the star or along
the equator, and a minimum in the center of the closed field lines. While the
magnetic field lines look qualitatively very similar in all cases, the current
distribution can be quite different.

Model A$_0$ has a spherically symmetric current distribution. The
corresponding magnetic field is stronger along the polar axis than along the
equator. It can be considered as the limiting configuration of
model C$_{c}$ for $c\rightarrow\infty$. Varying
the constant $c$, the spatial distribution of the current changes from being
spherically symmetric, $c\gg 1$, to being aligned with
the polar axis, $c\ll 1$ (see second and third panels in the top row of
Fig.\,\ref{fig_dipoles}).
The corresponding magnetic fields behave similarly, i.e. they align with the
polar axis for $c\ll 1$ (bottom row). 

By choosing a non-spherical current distribution, model A$_1$ (fourth column in
Fig.\,\ref{fig_dipoles}), the maximum current is found at the equator near the
outermost closed magnetic field line ($x\sim6\,$km). For this particular
configuration the current vanishes at the polar axis. The resulting magnetic
field is strongest in a central region oriented near the equator.
Model O (fifth column in Fig.\,\ref{fig_dipoles}) looks very similar,
but the current is somewhat more concentrated towards the equator, which leads
to a weaker 
magnetic field close to the polar axis. Finally, model F
 has a flat spectrum (see below) and is characterized by
two current maxima, one at the center and the other one at the equator at
$x\sim6\,$km. 

Due to their similar field configurations, the spectra of the corresponding
Alfv\'en oscillations are quite similar (see
Fig.\,\ref{fig_spectra}). Near the polar axis ($\chi=0\,$km) all spectra have a
turning point (U1) that can be either a maximum (A$_0$, C$_{c}$) or a
minimum (A$_1$, F, O). Here, $\chi$ is the radius where the magnetic field line
crosses the equatorial plane. The presence of turning
points in the spectra is of particular interest, since QPOs are expected at the
corresponding frequencies
\citep{Levin2007,Sotani2008,Cerda2009,Gabler2011letter,Gabler2012}.
The QPOs close to the polar axis are called {\it Upper QPOs}.

\subsubsection{Single Upper turning point (U1)}\label{sec_singleupper}

Model A$_0$, black line in Fig.\,\ref{fig_spectra}, was studied extensively
in \cite{Gabler2011letter, Gabler2012} and has a second turning point at
the closed field lines at $\chi\sim6.5\,$km, which causes the appearance of the
{\it Lower QPOs}
\citep{Cerda2009}. The spectra of the two models C$_{0.1}$ and C$_{10}$,
magenta and light blue lines in
Fig.\,\ref{fig_spectra}, are qualitatively similar. The main difference between
these two models is the gradient of the spectrum, which is
stronger for the smaller value of $c$, because of the difference in
rates at which the magnetic
field strength decreases with increasing $\chi$. For models where the electric
current is concentrated more along the polar axis, the magnetic field close to
the axis is stronger than for spherically symmetric currents, and it
decreases
faster with increasing $\chi$. The stronger the magnetic field is close to the
polar axis, the faster is the Alfv\'en speed in this region, i.e.
the frequencies near the polar axis increase, too. (cf. model
C$_{0.1}$). 
An abrupt change of the frequencies of the oscillations along neighbouring
field lines is expected to lead to faster phase-mixing, which contributes to
shorter lived QPOs \citep{Levin2007}. Observing differences in the lifetimes
of QPOs could in principle allow one to draw conclusions about the structure of
the magnetic field close to the polar axis. Unfortunately, due to the numerical
dissipation of our code we are not able to provide
reliable estimates for the differences of the QPO lifetime caused by
the different behaviour of the spectrum close to the polar axis.

\begin{figure}
\begin{center}	
\includegraphics[width=.47\textwidth]{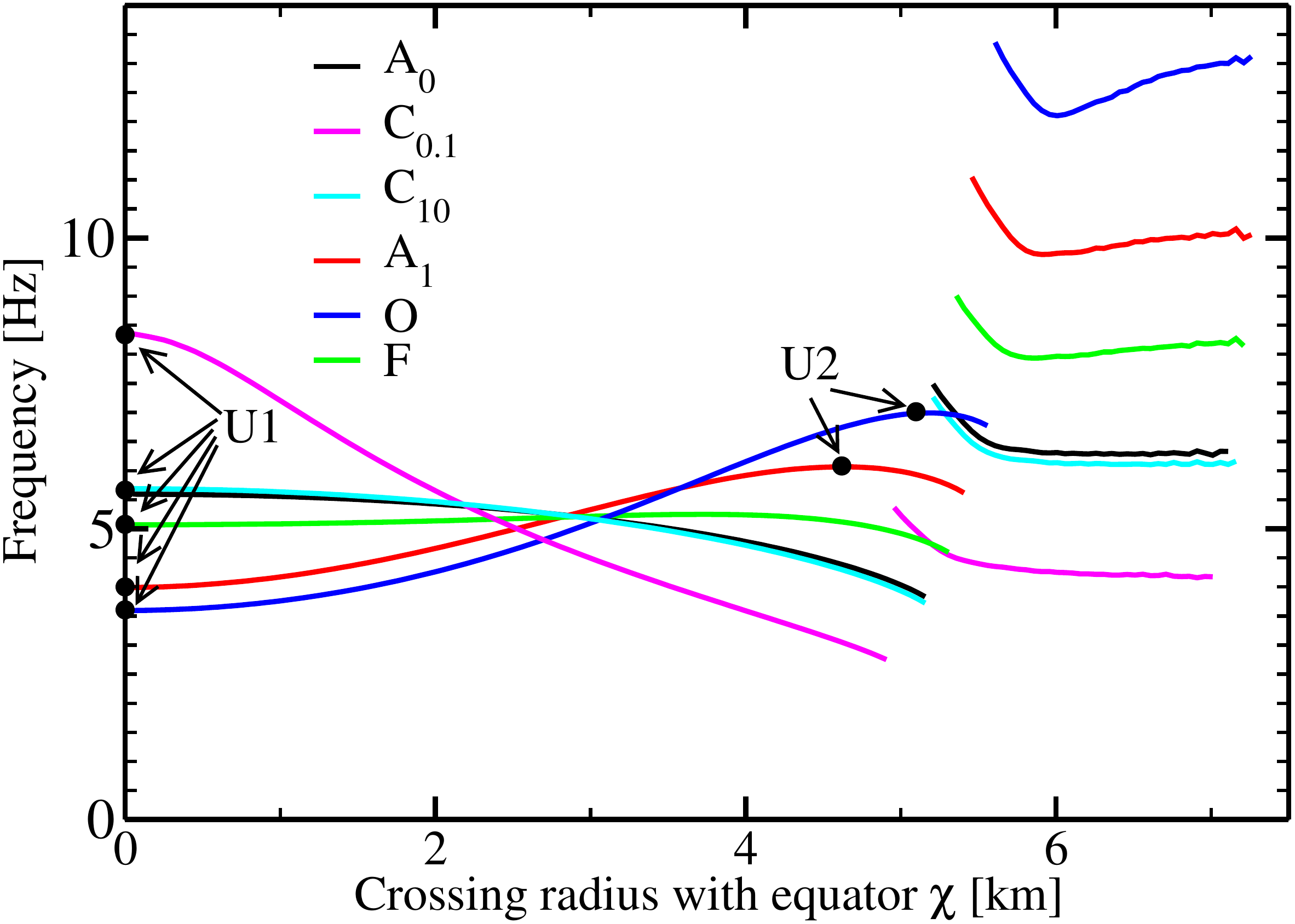}
\end{center}
\caption{ Alfv\'en spectra of the different magnetic field configurations
given in
Table\,\ref{tab_dipole_models}. The spectra are obtained with our
semi-analytic model at a surface-averaged magnetic field strength of
$10^{15}\,$G. The spectra for $\chi\lesssim5\,$km correspond to open field
lines, while those for $\chi\gtrsim5\,$km to closed ones. Upper turning points
(U1 and U2) are indicated black dots. The $x$-axis gives the point $\chi$ where
the
field line crosses the equatorial plane.}
\label{fig_spectra}
\end{figure}

The spectra for models C$_{c}$ have a minimum in the closed field line region
($\chi\gtrsim5\,$km) only for $c\gtrsim1$. Therefore, 
configurations with $c\lesssim1$ do not have lower QPOs as observed in
\cite{Gabler2012}. However, this may not have strong observational
consequences,
because closed field lines couple only indirectly through the crust to the
magnetosphere, i.e. we do not expect a strong modulation of
the emission occurring in the magnetosphere at the frequencies of the possible
lower QPOs.

The spectra of models A$_1$ and O are qualitatively different (red and blue
line in Fig.\,\ref{fig_spectra}). The turning point at the
polar axis is a minimum for both models, and there exists a second turning point
at $\chi\sim5\,$km, i.e. we expect a new family of QPOs (U2) besides the
previously
observed Upper (U), Lower (L) and Edge (E) QPOs.  Edge QPOs are
shorter-lived than turning-point QPOs. They are related to the edge of
the spectrum at $\chi\gtrsim5\,$km, where the outermost open field line crosses
the equator \citep[see][for
details]{Gabler2012}. 

\cite{Sotani2008, Cerda2009} and \cite{Gabler2012} interpreted the observed QPO
frequencies of 30, 92, and 150\,Hz in SGR 1806-20 as the
fundamental, the second and the fourth Alfv\'en overtones of the Upper QPOs, 
respectively. Similar arguments hold for SGR 1900+14 and the
observed frequencies of 28, 84, and 155Hz. The surface-averaged magnetic
field strength that matches the fundamental frequencies of both SGRs for the 
chosen current
configurations is listed in Table\,\ref{tab_freq_magstar}. It shows
that for models where the current has a maximum close to the polar axis
(C$_{0.1}$, C$_{10}$, and A$_0$), the surface magnetic field strength
necessary to match the observed $30\,$Hz in the SGR 1806-20 giant
flare is lower than for the other models (F, A$_1$, and O). 

\subsubsection{Two Upper turning points (U1 and U2)}\label{sec_doubleupper}

The presence of the second turning point in models A$_1$ and O allows 
for an other interpretation of the observed QPOs of SGR 1900+14: 
One could identify the fundamental QPO (U1$_0$) and its
second overtone (U1$_2$) with the observed frequencies of 28\,Hz and 84\,Hz, whereas
the observed frequencies of 53\,Hz and 155\,Hz could correspond to the fundamental
QPO at the second turning point (U2$_0$) and its
second overtone (U2$_2$) (located at $\chi\sim5\,$km)\footnote{ We use 
subscripts to indicate overtones of the fundamental QPO frequency.}.
Model O was constructed such as to fulfil this relation
between the observed frequencies. The corresponding model parameters 
(Table\,\ref{tab_dipole_models}) are not very peculiar, i.e.
given the current theoretical uncertainties model O
is as valid as any other model we could have constructed. 

\begin{table}
\begin{center}
\begin{tabular}{c | c c c c}
Model&\multicolumn{2}{c}{turning point polar
axis (U1)}&\multicolumn{2}{c}{second
turning point (U2)}\\
& $B_{15}$ & $B_{15}$& $B_{15}$& $B_{15}$\\
&$f_0=28\,$Hz&$f_0=30\,$Hz& $f_0=28\,$Hz&$f_0=30\,$Hz\\
\hline
A$_0$&4.9&5.3\\
C$_{0.1}$&3.4&3.6\\
C$_{10}$&4.9&5.3\\
A$_1$&6.8&7.4&4.7&5.0\\
O &7.8&8.3&4.1&4.3\\ 
F &5.5&5.9\\\hline
$R_4$&4.1&4.4&3.4&3.6\\
$R_5$&5.4&5.9&3.0&3.2\\
$R_6$&7.4&7.9\\
$R_7$&10.3&11.0\\
$R_8$&14.3&15.2\\ \hline
\end{tabular}
\caption{ Surface-averaged magnetic field strength needed to match a turning
 point of the spectrum to $f_0=28\,$Hz and $f_0=30\,$Hz, respectively.
 The second and third column match the frequency
of the turning point at the polar axis (U1), while the last two columns
match the frequency of the second turning point (U2) in
the spectrum, if present (A$_1$, O, $R_4$, and
$R_5$).
}\label{tab_freq_magstar}
\end{center}
\end{table}
\begin{figure}
\begin{center}	
\includegraphics[width=.475\textwidth]{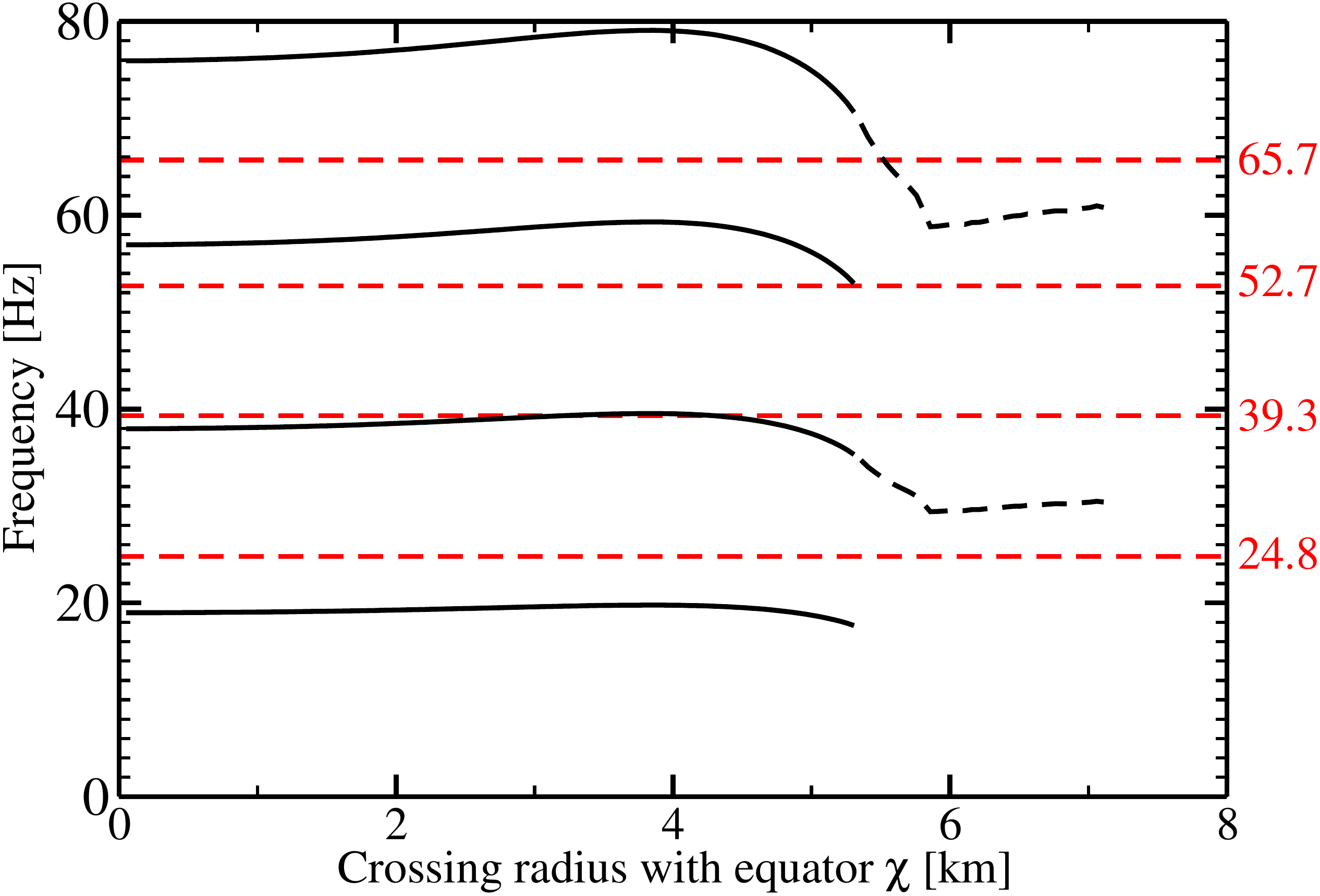}
\end{center}
\caption{ Alfv\'en spectrum of the magnetic field configuration F which
leads to a very
flat spectrum at $B_{15}=3.7$. Red dashed lines indicate the frequencies of the
crustal shear modes. The $x$-axis gives the point $\chi$ where the
field line crosses the equatorial plane.}
\label{fig_flat_spectra}
\end{figure}
\begin{figure}
\begin{center}	
\includegraphics[width=.475\textwidth]{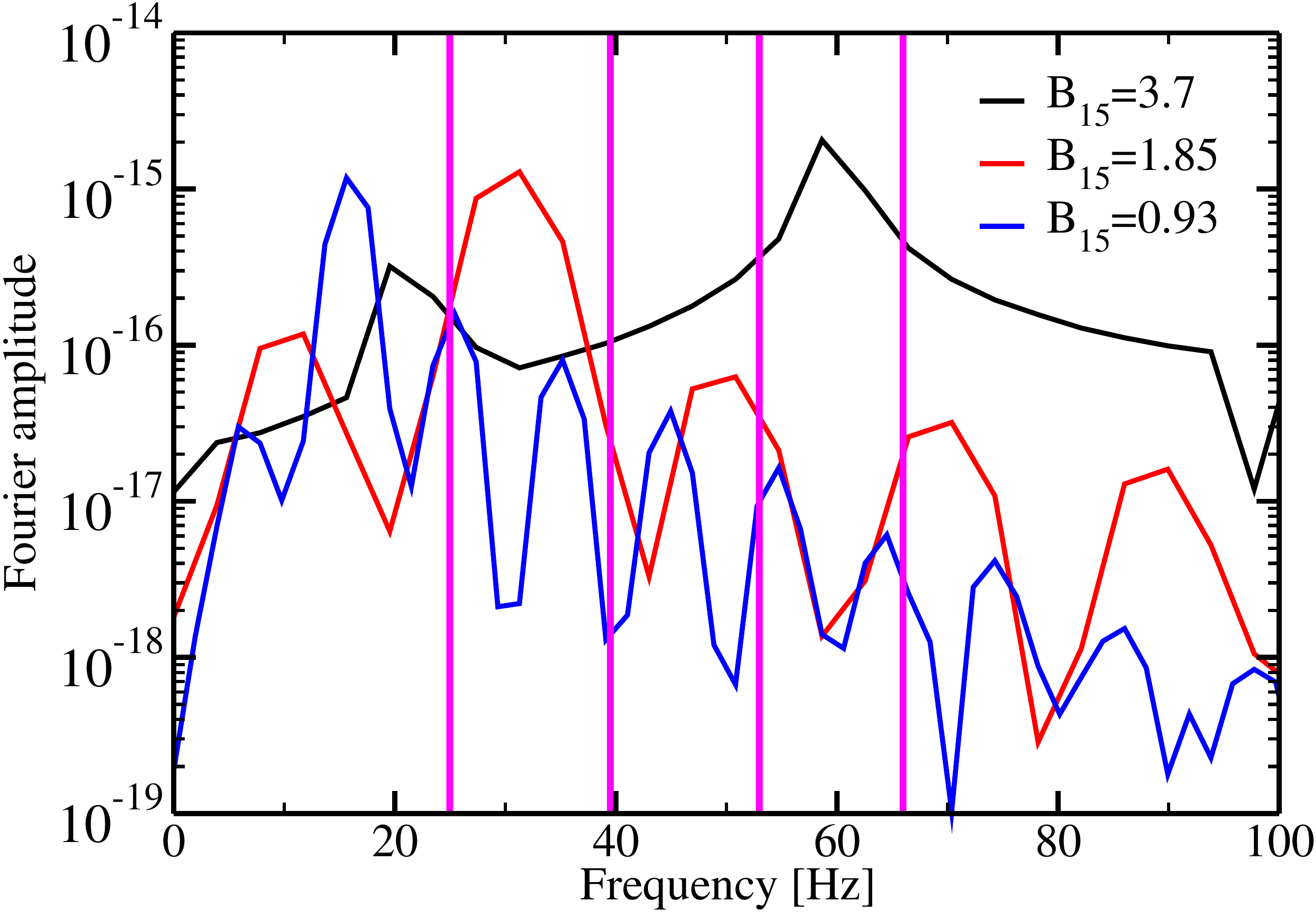}
\end{center}
\caption{Fourier analysis of the oscillations at $B_{15}=3.7$,
$1.85$, and $0.93$. Magenta vertical lines
indicate the frequencies of the first four crustal shear modes, whose precise values are given in Fig.~\ref{fig_flat_spectra}.}
\label{fig_flat_FFT}
\end{figure}

\subsubsection{ Nearly flat spectra: no shear modes inside continuum
gaps}\label{sec_gaps}

We choose the parameters of model F to obtain a very flat spectrum
which is expected to produce very long-lasting QPOs in a large volume of the
star. 
This happens because for almost all open field lines the QPOs have a very
similar
frequency. Additionally, the gap between successive overtones in the spectrum is
maximized, which may allow for crustal shear modes surviving in these gaps
\citep{Colaiuda2011, vanHoven2012} instead of being absorbed into the
Alfv\'en continuum of the core as in \cite{Gabler2012}. 
Hence, we have computed the spectra of the fundamental 
Alfv\'en oscillation and its first three overtones at $B_{15}=3.7$. These
spectra are shown in Fig.\,\ref{fig_flat_spectra} together with the frequencies
of the crustal shear modes (horizontal red dashed lines). The $l=2$ and $l=5$
shear modes  with 24.8\,Hz and 65.7\,Hz, respectively, clearly fall between the
frequencies of Alfv\'en overtones, while the frequencies for the $l=3$ and
$l=4$ modes lie just on top of the maximum or at the edge.
To search for the crustal shear modes which may have been survived, we performed
simulations covering a period of 0.25\,s, calculated the overlap
integrals\footnote{Overlap integrals 
are global measures of how strong a given shear mode is excited. For details we
refer to \cite{Gabler2012}.} with the pure crustal shear modes, and
Fourier-analyzed them.
The resulting Fourier transform for the $l=2$ crustal shear mode at
$B_{15}=3.7$, $1.85$, and $0.93$ is given in
Fig.\,\ref{fig_flat_FFT}. At $B_{15}=3.7$ we find strong signals at
the frequencies $f\sim20\,$, $60$, and $100\,$Hz. 
These are the frequencies expected from the semi-analytic
model for the fundamental QPO and its second and fourth overtone
(see Fig.\,\ref{fig_flat_spectra}). 
The frequencies of the overtones obey integer relations and do not behave as
those of crustal shear modes
which are proportional to $f_l\sim \sqrt{(l-1)(l+2)} f_0$. {\it None of the
frequencies of the first three shear modes indicated by the magenta vertical
lines in Fig.\,\ref{fig_flat_FFT} coincides with an observed frequency of a QPO
at $B_{15}=3.7$.} 
Fig.\,\ref{fig_flat_FFT} also shows that all observed
QPOs scale directly with the magnetic field: the frequencies at $B_{15}=3.7$ are
twice as large as the corresponding frequencies at
$B_{15}=1.85$ and are four times as large as at $B_{15}=0.93$.
The matching of the frequency of the Alfv\'en overtone at $f\sim25\,$Hz and at
$B_{15}=0.93$ with that of the $l=2$ shear mode is pure coincidence.

Qualitatively the same result holds when analyzing the overlap
integral for the $l=3$ crustal shear mode. In this case all QPOs where the
velocity field is symmetric with respect to the equatorial plane are discovered
and, again, {\it no signal is present at the crustal shear modes}. The
analysis with the overlap integrals is limited to oscillations having the same
equatorial symmetry as the corresponding shear mode, i.e. the overlap integral
is zero for the $l=2$ ($l=3$) mode and an oscillation of odd
(even) symmetry with respect to the equatorial plane.

\subsubsection{Ring current}\label{sec_ring_current}
\begin{figure}
\begin{center}	
\includegraphics[width=.475\textwidth]{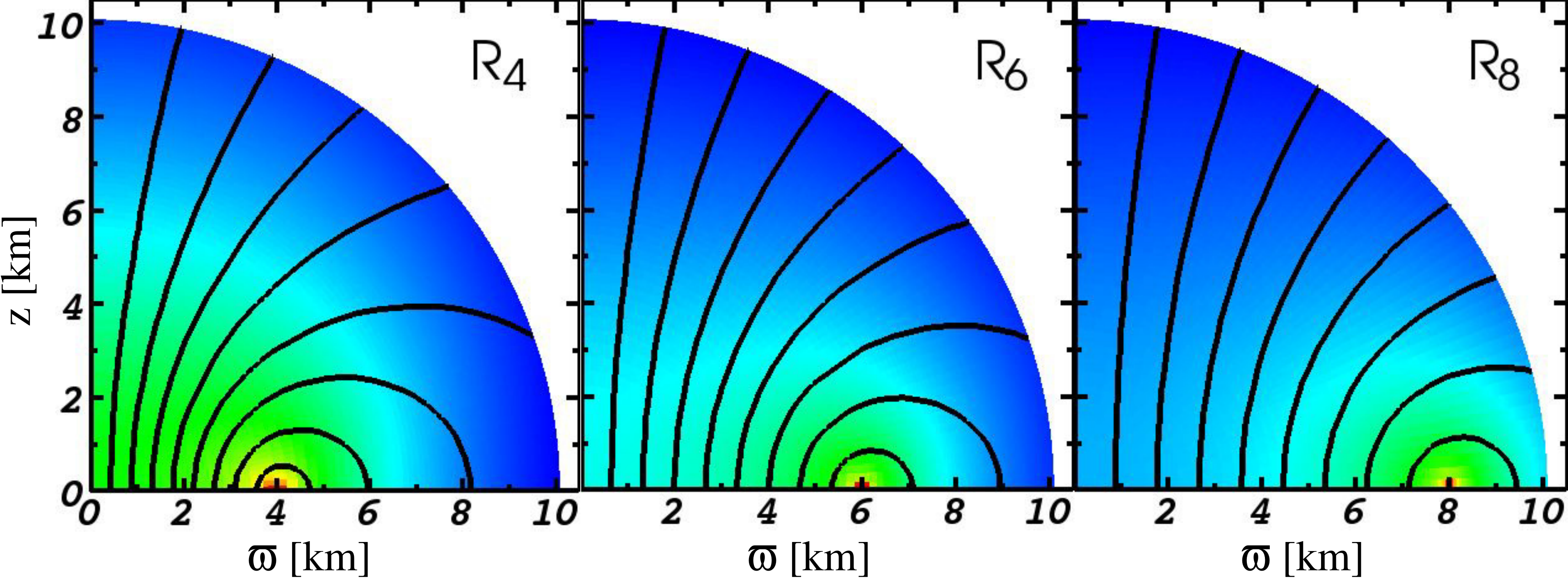}
\end{center}
\caption{ Magnetic field configuration for a ring current at a given radius
as indicated in the panels. The colour
scale ranges from blue (minimum) to red (maximum), and gives the logarithm of
the magnetic field strength.}
\label{fig_ring}
\end{figure}

Next we consider the extreme case of a magnetic field generated
by a circular current loop inside the star, whose 
analytical Newtonian solution can be found in~\cite{Jackson1998}.
A ring current can be seen as a limiting case of model A$_1$ where the
region of the current is shrunk to a ring. Placing the current at
different radii
we can construct different magnetic field configurations inside the neutron star
(Fig.\,\ref{fig_ring}).
\begin{figure}
\begin{center}
\includegraphics[width=.47\textwidth]{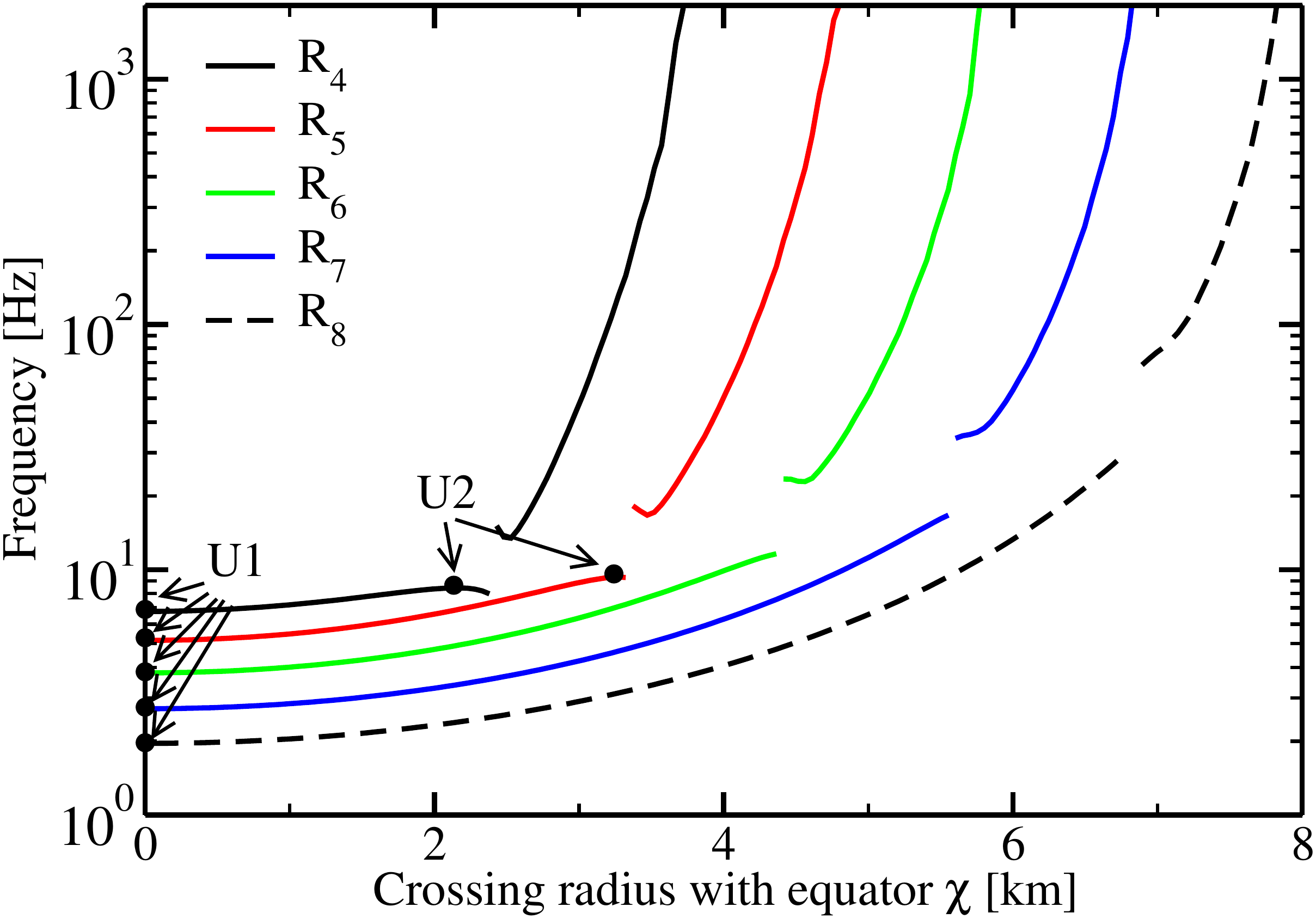}
\end{center}
\caption{ Alfv\'en spectra of magnetic field configurations obtained from
ring currents $R_\varpi$ at different radii $\varpi$. The spectra are
calculated with our semi-analytic model for a surface-averaged magnetic field
strength of $B_{15}=1$. The $x$-axis gives the point $\chi$  where the
field line crosses the equatorial plane.}
\label{fig_spect_ring}
\end{figure}
The effects on the Alfv\'en oscillations are illustrated in
Fig.\,\ref{fig_spect_ring}, which displays the spectra for models $R_\varpi$
where the ring currents are located at $\varpi=4$, $5$, $6$, $7$, and
$8\,$km from the center. In all cases the spectra have a minimum at the polar
axis $\chi=0\,$km (U1), and the
frequency of the open field lines increases with increasing $\chi$. 
Similar to model A$_1$ there is a second maximum at $\chi \sim
2.0\,$km and $\chi \sim 3.0\,$km for models $R_4$ and $R_5$, respectively. This
second
turning point may give rise to a new family of QPOs (U2) as discussed for models
A$_1$ and O in the previous subsection.

The spectra associated with the closed field lines (right part of
Fig.\,\ref{fig_spect_ring}) show a strongly increasing frequency with
increasing $\chi$ until  the location of the ring current is reached. There are
additional turning points for models $R_4$, $R_5$, and $R_6$ at $\chi \sim
2.5\,$km, $\chi \sim 3.5\,$km and $\chi \sim 4.5\,$km, 
respectively. However, the steepness of the spectra limits the existence of
long-lived lower QPOs for these models, and for models
$R_7$, and $R_8$ we do not expect lower QPOs at all.

\subsection{Mixed poloidal-toroidal field (type I) }\label{sec_field_LoreneII}
 
We turn now to models that have a mixed poloidal-toroidal field. 
In this case,
the system of equations presented in Sec. \ref{subsec_mhd} is no longer valid,
as the toroidal component will couple axial and polar oscillations. However, 
since the toroidal component of our models is limited to the region of closed field
lines inside the star, axial and polar oscillations still decouple at the
linear level in the region of the open field lines. 
The coupling in the closed field line region will excite polar
oscillations in the whole star. Fortunately, these do not influence the
continuum at the linear level and we can still use our semi-analytic model to
estimate the frequencies of the Alfv\'en oscillation in the region of open field
lines.

Our models are obtained setting $a_0=1$ in Eq.\,(\ref{eq_poloidal_models}) and
Eq.~(\ref{eq_mixed_models}) with different values for $b_0$.
The surface-averaged magnetic
field strengths and the maximum poloidal and toroidal field strength for the
respective configurations are given in Table\,\ref{tab_pol_tor}. The two
components have similar strength at $b_0\sim10$.
\begin{table}
\begin{center}
\begin{tabular}{c  c |  c c c}
Model & $b_0$&\multicolumn{3}{c}{magnetic field strength [$10^{12}\,$G]}\\
&&surface-averaged & max. poloidal&max. toroidal\\\hline
B$_0$&0&3.5&19&0\\
B$_1$&1&3.6&20&3.5\\
B$_2$&2&4.0&21&7.1\\
B$_5$&5&5.6&24&18\\
B$_{10}$&10&7.6&27&33\\
B$_{20}$&20&8.9&31&54
\end{tabular}
\caption{Surface-averaged, maximum poloidal and toroidal magnetic
field strength for mixed poloidal-toroidal configurations with parameters
$a_0=1$ and $b_0$ as indicated.}\label{tab_pol_tor}
\end{center}
\end{table}
\begin{figure}
\begin{center}
\includegraphics[width=.47\textwidth]{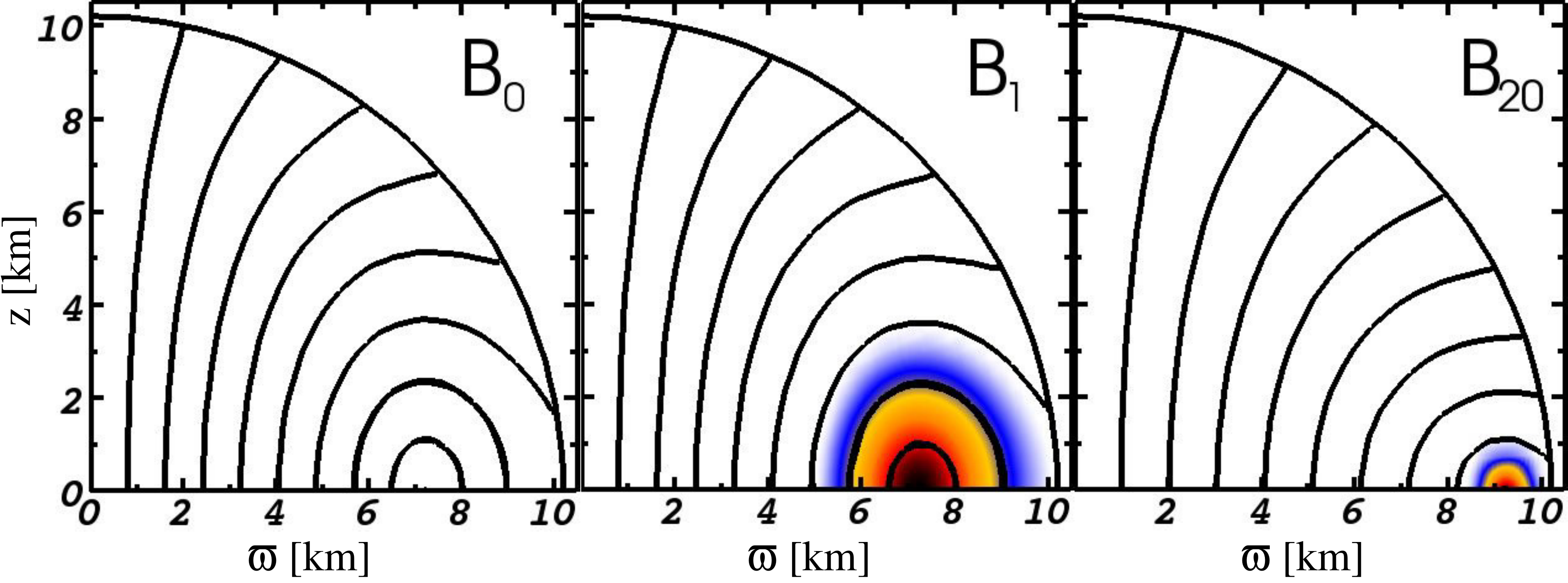}
\end{center}
\caption{Magnetic field lines of different mixed poloidal-toroidal models.
The colour scale gives the strength of the toroidal magnetic field and ranges
from white-blue (minimum) to red-black (maximum). The field of model B$_0$ is
purely poloidal.}
\label{fig_pol_tor}
\end{figure}
The field lines
and the toroidal magnetic field strength of selected models are plotted in
Fig.\,\ref{fig_pol_tor}, which shows that the area occupied by the closed
field lines is largest in the
case of zero toroidal field (model B$_0$). With increasing toroidal field this
region shrinks
and shifts towards the surface, see e.g. \cite{Lander2012a}. Additionally, the
strength of the poloidal component increases at the closed field lines and at
an limited region of the open field lines around the closed ones.

In Fig.\,\ref{fig_spect_pol_tor} we plot the spectra of the open
field lines of our mixed field models. For zero
toroidal field the Alfv\'en spectrum is similar to the black line in 
Fig.\,\ref{fig_spectra}, but scaled at different frequencies. 
There is a maximum at the polar axis ($\chi=0\,$km) and an
edge at $\chi\sim5\,$km. With increasing values of $b_0$  not only 
the toroidal field increases but also the poloidal one. This increase is stronger near
the closed field lines, such that the frequencies at large $\chi$ ($\sim5\,$km)
increase faster than closer to the polar axis, leading to a flatter spectrum
in the case of model B$_2$  (green line in the figure). With an even stronger
toroidal component, the turning point at the polar axis changes from a maximum
to a minimum, and the Alfv\'en oscillations near the closed field lines
($\chi\sim 6$ and $7\,$km for models B$_5$ and B$_{20}$, respectively) show the
highest frequencies. The larger $b_0$ is, the steeper  the gradient of the
spectrum becomes
at the edge. In this case possible QPOs at the edge would last for shorter
timescales than for model B$_0$ where the gradient of the spectrum is moderate.

\begin{figure}
\begin{center}
\includegraphics[width=.47\textwidth]{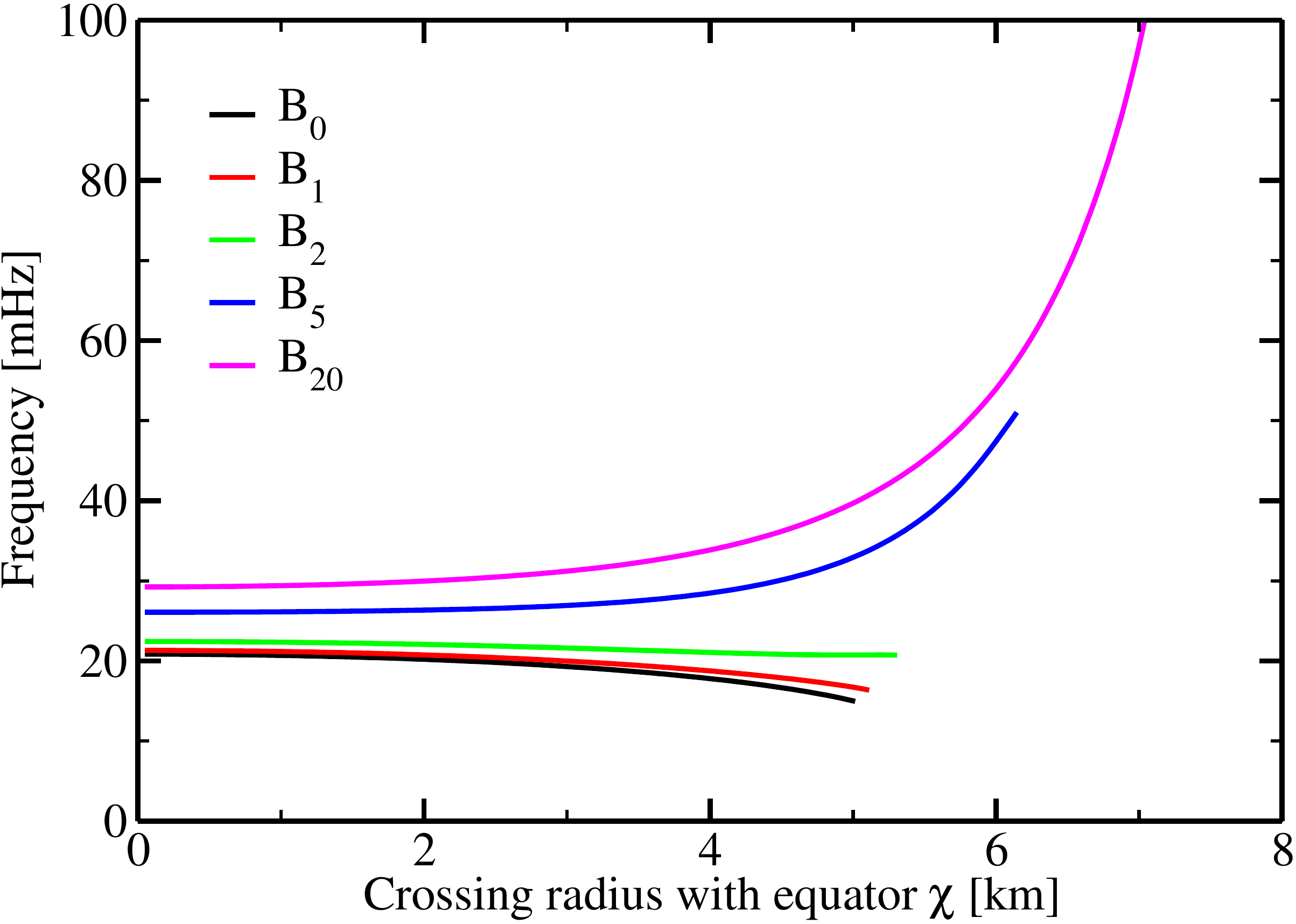}
\end{center}
\caption{Alfv\'en spectra of torsional oscillations along
the open field lines for mixed poloidal-toroidal models B$_{x}$ obtained 
with our semi-analytic method. See Table\,\ref{tab_pol_tor} for model details.
The $x$-axis gives the point $\chi$ where the field line crosses the equatorial
plane.}
\label{fig_spect_pol_tor}
\end{figure}

\section{Mixed dipole-quadrupole-like poloidal configurations}\label{sec_quadru}

To obtain configurations with a quadrupole-like contribution
we only consider for simplicity purely poloidal magnetic fields, i.e. type II
magnetic fields, by setting $\tilde{H}(A_\varphi)$ equal to zero.  Concerning
the choice of parameters for the function $M(A_\varphi)$,
only configurations that allow for a change of sign of ${\mathcal
J}^\varphi$ across the equator and the
corresponding initial data can be considered. In this section we
choose $a_1\neq0$, while all other parameters are set to zero.

As we have mentioned in Section\,\ref{sec_bfield_magnetar} we construct mixed
dipole-quadrupole-like magnetic field configurations by linearly superimposing
both
currents, which are obtained independently. For the purely quadrupole-like
(dipole-like) configuration, Q (D), $a_1=18.1$
($a_0=11.66$) and the averaged surface magnetic field strength is
$B=6.0\times10^{13}\,$G
($B=4.4\times10^{13}\,$G). When constructing the mixed configurations we keep
the dipole-like field strength fixed and add a quadrupole-like
component rescaled by $Q/D$ times the reference field. 

The field lines of selected mixed configurations are displayed in
Fig.\,\ref{fig_quad}. The configuration with $Q/D=10$ is predominantly
quadrupole-like, while $Q/D=0.1$ gives rise to a slightly deformed dipole-like
configuration. When the amplitudes of both components ($Q/D= 1$) are comparable,
the field
is very distorted, and has no symmetry with respect to the equatorial plane.

We emphasize that we study the mixed dipole-quadrupole-like configurations {\it
without including
a shear modulus in the crust}, in order to isolate the effect that the inclusion of 
a quadrupole-like component has on the continuum of torsional oscillations.

\label{sec_quad}
\begin{figure}
\begin{center}
\includegraphics[width=.47\textwidth]{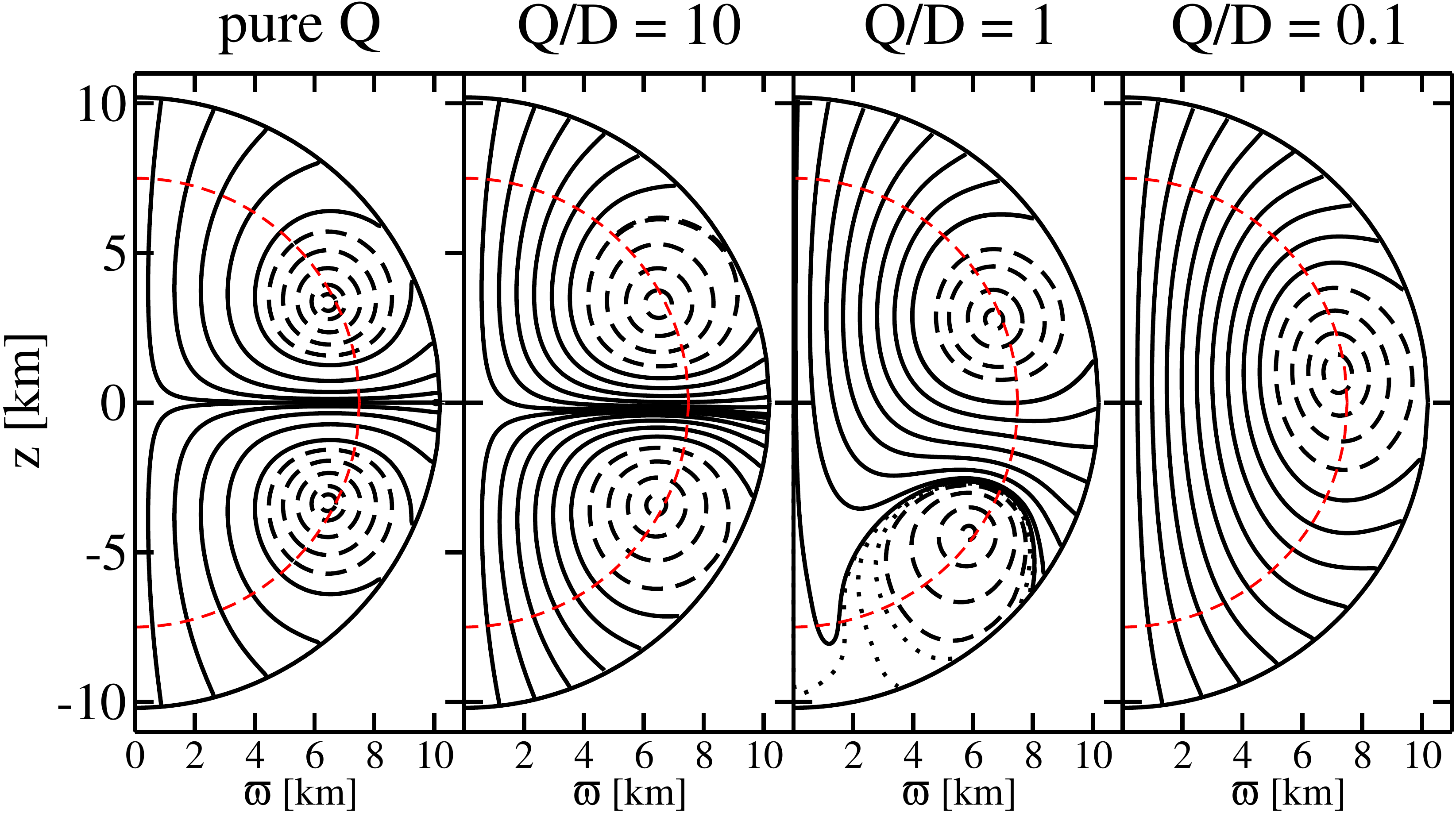}
\end{center}
\caption{Field lines of different mixed dipole-quadrupole-like (D-Q)
configurations, the left
panel showing the purely quadrupole-like case. The title of each panel indicates
the Q-to-D 
component ratio. The red, dashed lines indicate the path along which the spectra of Fig.\,\ref{fig_spect_quad} have been obtained.}
\label{fig_quad}
\end{figure}
\subsection{Alfv\'en spectra using the semi-analytic model}
\begin{figure*}
\begin{center}
\includegraphics[width=.9\textwidth]{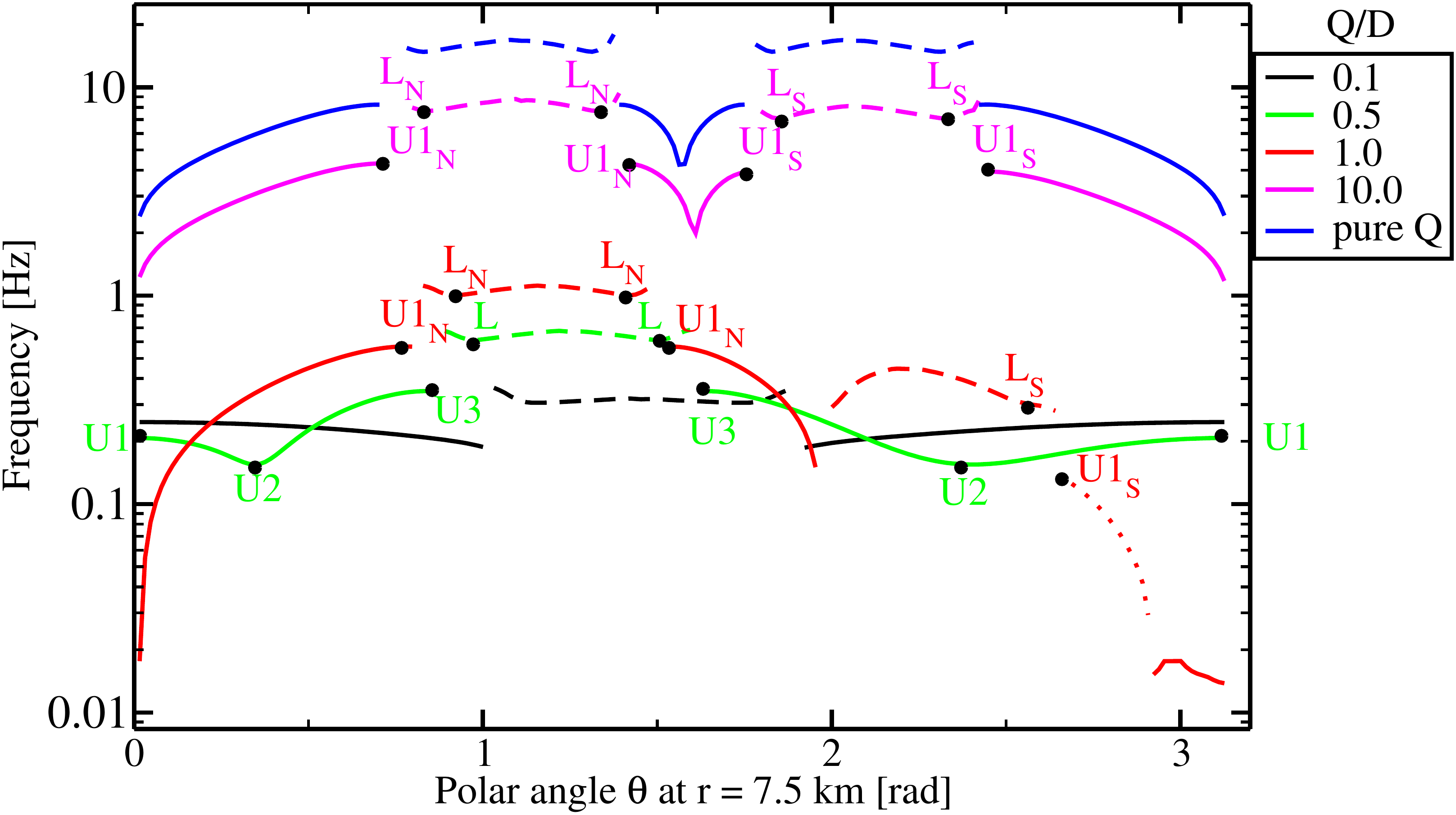}
\end{center}
\caption{ Alfv\'en spectra of different mixed
configurations. Different line styles indicate regions of different types of
field lines corresponding to Fig.\,\ref{fig_quad}. Solid lines correspond to
open field lines, dashed lines to closed ones and the dotted line
corresponds to open field lines only existing in the southern hemisphere for
$Q/D=1.0$. Selected turning points are indicated with black dots and labelled
with $U$ for Upper QPOs and $L$ for Lower QPOs, respectively. The subscripts N
and S stand for the hemisphere where the QPO appears, the northern
$0<\theta<\pi/2$ (N) and the
southern $\pi/2<\theta<\pi$ (S).
The field strength of the dipole-like component is
$4.4\times10^{13}\,$G, and the quadrupole-like component varies from
$0.6\times10^{13}\,$G at $Q/D=0.1$ to $6.0\times10^{14}\,$G at $Q/D=10.0$,
respectively.
The pure quadrupole-like configuration is shown for $B=1.2\times10^{15}\,$G.}
\label{fig_spect_quad}
\end{figure*}
In Fig.\,\ref{fig_spect_quad} we plot the spectra of different combinations
of Q and D. The $x$-axis gives the polar angle $\theta$ at which the field line
crosses a circle with radius $7.5\,$km indicated with red, dashed lines in
Fig.\,\ref{fig_quad}.
This choice is motivated by capturing all field lines of the spectrum in one
single parameterization . Fortunately, the center of the region(s) of closed
field lines lies almost on this circle for all configurations studied here.

For $Q/D=0.1$ (black lines in Fig.\,\ref{fig_spect_quad}) the spectrum looks
similar to the pure dipole-like case $A_0$. Qualitatively, we can identify the 
end of the line at $\chi\sim5.2$ km in Fig.\,\ref{fig_spectra} with $\theta\sim1.0$ rad
in Fig.\,\ref{fig_spect_quad}. Similarly, the 
local maximum at $\chi\sim7.0$ km in one figure can be identified with the
local 
maximum $\theta\sim1.5$ rad in the other figure.
We note the asymmetry between
the two hemispheres ($\theta<\pi/2$ and $\theta>\pi/2$) due to the presence of
the
quadrupole-like component. There is a maximum at the polar axis in the
northern and the southern hemispheres for the open field lines (solid lines in
the figure) and a minimum 
inside the region of closed field lines (dashed lines in the figure) at
$\theta\sim1.1$. The latter field lines also form the minimum at
$\theta\sim1.7$.
Additionally, there is another maximum at $\theta\sim1.5$ that is associated
with
the closed field line which collapses into one point. Since this line
has vanishing length, we do not expect to see a QPO at this location. 

When the quadrupole-like component is half as strong as the dipole-like one, 
model $Q/D=0.5$ (green lines in Fig.\,\ref{fig_spect_quad}), there appear two
new
extrema in the spectrum of the open field lines at $\theta\sim0.4$ (U2) and
$\theta\sim0.8$ (U3),respectively, which could produce QPOs. Additionally, the
spectrum becomes
fairly asymmetric and the part due to the closed field lines (dashed lines in the figure) lies 
almost completely in the northern hemisphere $\theta<\pi/2$. At equal amplitudes
of both magnetic fields, model $Q/D=1.0$ (red lines in
Fig.\,\ref{fig_spect_quad}), the part of the spectrum due to the open
field lines simplifies: the turning points at $\theta=0.0$ and
$\theta\sim0.5$ disappear, and only the one maximum at $\theta \sim0.75$
(U1$_N$)
remains. The closed field lines part of the spectrum (dashed lines) in the
northern hemisphere 
is now limited to the interval $0.8\lesssim\theta\lesssim1.5$., and for
$2.0\lesssim\theta\lesssim2.5$ there appears a new family of
closed field
lines disconnected from the previous ones at smaller $\theta$. The new closed lines have
lower fundamental Alfv\'en frequencies, because the magnetic field in this area
of the quadrupole-like component cancels partly with that of the dipole-like
component, i.e. the resulting field is weaker than in the northern
hemisphere. 
Additionally, there is a new family of open field lines (dotted red
line in the figure) surrounding the closed lines at $\theta>\pi/2$, which we
only observe for
this particular combination $Q/D=1.0$. They do not have any turning point,
i.e. we expect at most very weak QPOs at the edges of the spectrum.

For increasing values of $Q/D$ the frequencies increase
for field lines with $\theta>0.5$, because of the
high magnetic field strength of the quadrupole-like component in this region.
For the model with $Q/D=10.0$ (magenta lines in Fig.\,\ref{fig_spect_quad}) the
symmetric spectrum of a pure
quadrupole-like configuration
is approached (see the blue lines in Fig.\,\ref{fig_spect_quad}). For $Q/D=10.0$
there is one
turning point for the open field lines at $\theta\sim0.7$ (U1$_N$) that is
formed by the same field lines leading to another turning point at
$\theta\sim1.4$.
They are also related to the two turning points (U1$_S$) at $\theta\sim1.7$ and
$\theta\sim 2.4$, respectively, caused by the mirrored field lines in the
opposite hemisphere. The
form of the spectrum associated with the closed field
lines almost does not change in the northern hemisphere $\theta<\pi/2$ compared
to $Q/D=1.0$, while the one in the southern hemisphere becomes symmetric
to the one in the northern hemisphere. With increasing $Q/D$ ratio, also the
absolute values of the frequencies increase.

\begin{figure*}
\begin{center}
\includegraphics[width=.94\textwidth]{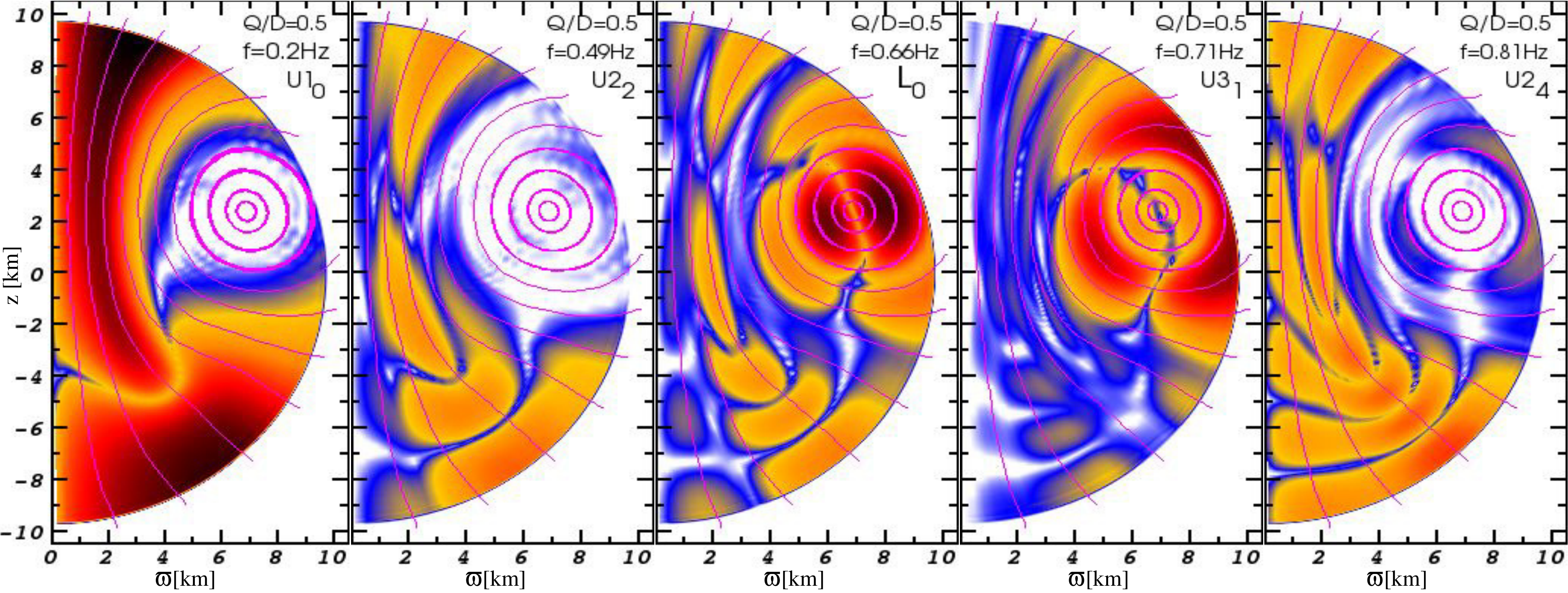}
\includegraphics[width=.94\textwidth]{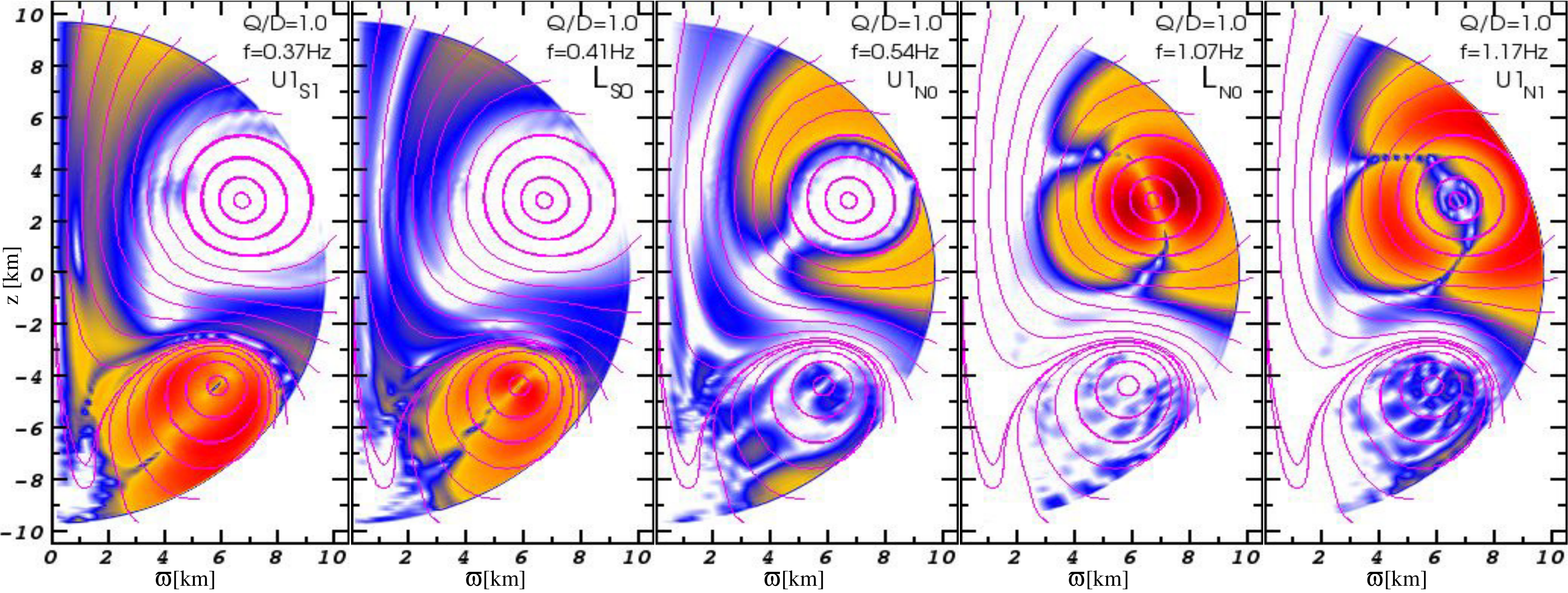}
\includegraphics[width=.94\textwidth]{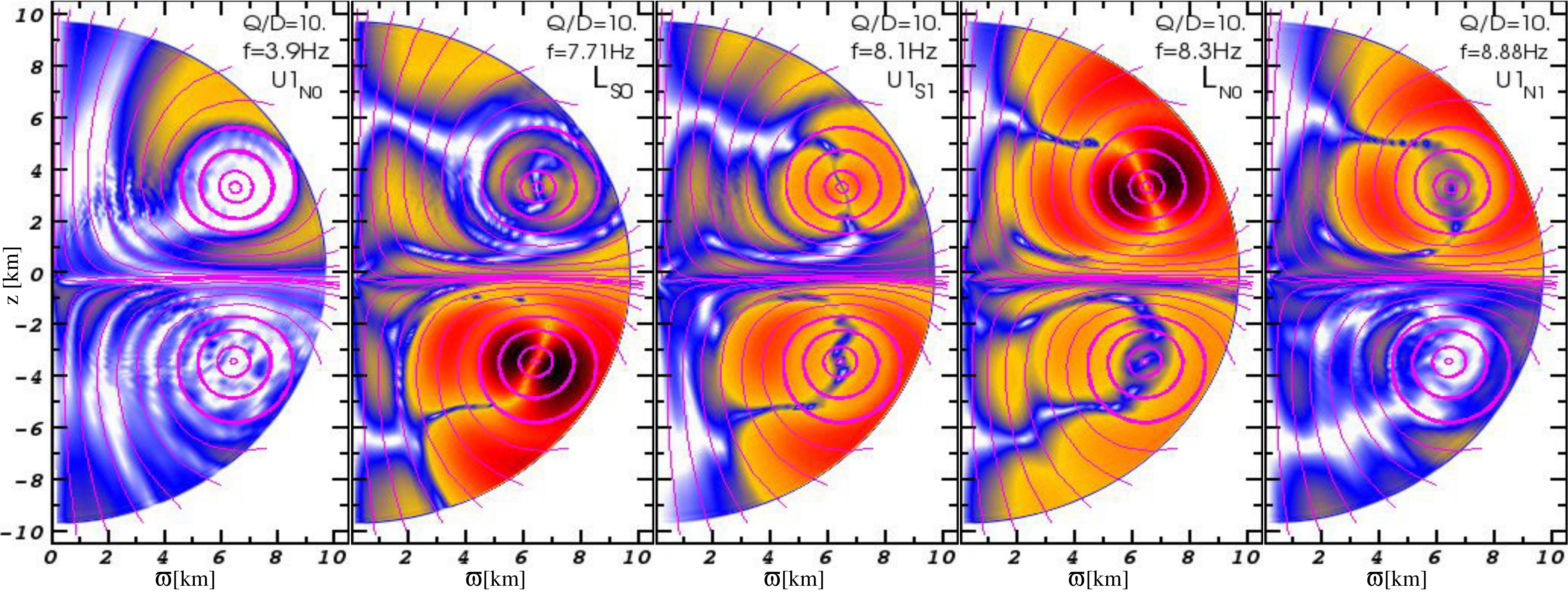}
\end{center}
\caption{Fourier transforms of the Alfv\'en QPOs for configurations $Q/D=0.5$ (top),
$Q/D=1$ (middle), and $Q/D=10$ (bottom), respectively.
The plotted QPOs are: $U1_0$, $U2_2$, $L_0$, $U3_1$, and $U2_4$ (top), 
$U1_{S1}$, $L_{S0}$, $U1_{N0}$, $L_{N0}$, and $U1_{N1}$ (middle)
$U1_{N0}$, $L_{S0}$, $U1_{S1}$, $L_{N0}$, and $U1_{N1}$ (bottom).
The subscript indicates the number of the overtone, with $0$ indicating the
fundamental QPO. In all panels the colour scale ranges from white-blue (minimum)
to red-black (maximum) and gives the logarithm of the Fourier amplitude of the QPOs 
at the frequency denoted in the upper right of each panel.
}
\label{FFT_mix}
\end{figure*}

\subsection{ Alfv\'en QPOs using simulations}

For the study of the time evolution of the Alfv\'en oscillations of the mixed
dipole-quadrupole-like configurations we use the GRMHD code described
in Sec. \ref{subsec_mhd}.
The Fourier transforms of the strongest QPOs during the simulations for the configurations 
$Q/D=0.5$, $1.0$, and $10.0$ are shown in the top, middle, and
bottom rows of 
Fig.\,\ref{FFT_mix}, respectively.  We find oscillations at the frequencies
predicted by the spectrum obtained with the semi-analytic model. 
The particular frequencies and names of the QPOs displayed in
Fig.\,\ref{FFT_mix} are indicated in the upper right of each panel. 
We selected those QPOs which are best suited for the
visualization considering different effects. For example, the
particular initial perturbation applied to the equilibrium model may have
excited some QPOs more than others. Some QPOs may have similar frequencies,
i.e.  their signals may overlap in the corresponding Fourier transform,
which has a finite resolution. As expected, the spatial structure of the QPOs
shown in the
figure follows the structure of the magnetic field lines.  For the $Q/D=0.5$
(top row)
the maximum amplitude of the Fourier transform for the $U1_0$ QPO (first column)
is located near the polar axis, for $U3_1$ (fourth column) it is at the open
field lines nearest to the closed field lines, while for $U2_2$ and $U2_4$
(second and fifth columns) 
the maximum amplitudes are found in between these two regions.
The oscillation of $L_0$ (third column) is limited to the closed field lines.
In the third column we can also see three overtones of the different Upper
QPOs: $U1_2$ (three nodes near the polar axis), $U2_3$ (four nodes), and
$U3_1$ (two nodes along the open field lines close to the closed ones), which
all have approximately the same frequency.

The QPOs of the configuration with $Q/D=1.0$ are
plotted in the middle row of Fig.\,\ref{FFT_mix} and are labelled $U1_{S1}$,
$L_{S0}$, $U1_{N0}$, $L_{N0}$, and $U1_{N1}$ (from left to right, respectively).
The additional
subscripts $N$ and $S$ indicate the northern or southern hemisphere where
different QPOs are localized. $U1_{S0}$ and $L_{S0}$ are the fundamental
oscillations of the open and closed field lines in the southern hemisphere,
while $U1_{N0}$, $L_{N0}$, and $U1_{N1}$ are limited to the northern one. 

For $Q/D=10$ (bottom row) both hemispheres approach a similar
magnetic field configuration, which results in an almost symmetric distribution of the
oscillations. In most of the panels (2-4) of this row we see a
strong Fourier amplitude in both hemispheres. This is expected because the
frequencies of the corresponding QPOs in the northern and southern hemispheres
are almost equal. However, there are still some small differences in the
frequencies, e.g. between $U1_{S1}$ ($f\sim8.1\,$Hz) and $U1_{N1}$
($f\sim8.9\,$Hz), or between $L_{S0}$ ($f\sim7.7\,$Hz) and  $L_{N0}$
($f\sim8.3\,$Hz), respectively. 

\begin{figure*}
\begin{center}	
 \includegraphics[width=.75\textwidth]{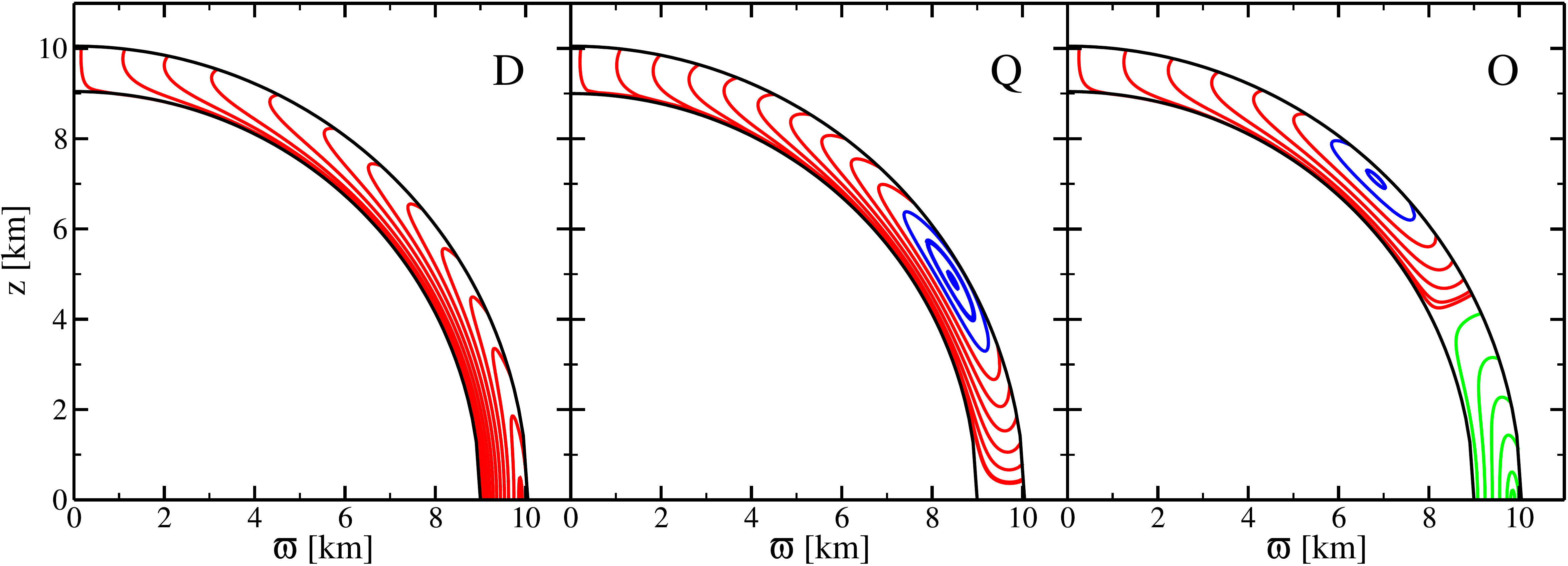}
\includegraphics[width=.78\textwidth]{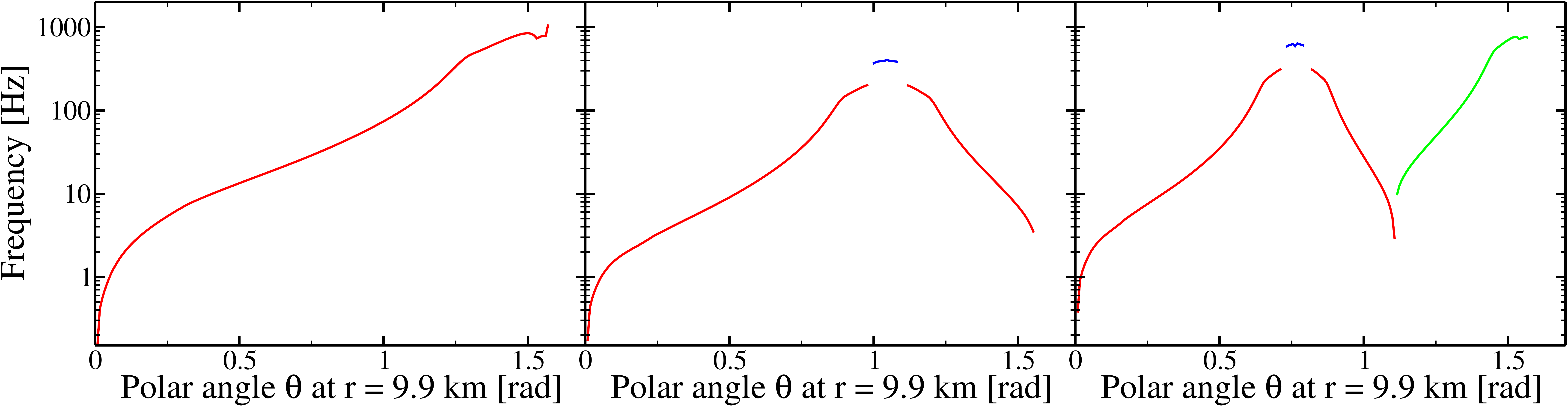}
\end{center}
\caption{{\it Upper panels}: Magnetic field configurations for models matched
to an exterior dipole (left), quadrupole (middle), and octupole
(right), respectively. {\it Lower panels}: Alfv\'en spectra of the
magnetic field
configurations given in the corresponding upper panel obtained with the
semi-analytic model
for an averaged surface magnetic field of $B\sim10^{14}\,$G. Corresponding 
field lines in the upper and lower panels have the same colour.
}
\label{fig_crustfield}
\end{figure*}

\section{Magnetic field configurations confined to
the crust}\label{sec_crustfield}
We now turn to the study of magnetic field configurations that are confined to
the crustal region of the neutron star.  In such cases the crustal shear modes 
cannot be absorbed by the core. Our choice is motivated by the possible presence 
of superconducting protons in the core of the neutron
star \citep{Page2011,Shternin2011} which could expel the magnetic flux due to
the Meissner-Ochsenfeld effect. From theoretical studies \citep{Baym1969} one
would expect the protons to form a superconductor of type II,  the magnetic
field being able to penetrate the superconducting region in form of
flux tubes. However,
due to the uncertainty of the EoS at supranuclear densities, the opposite situation 
cannot be ruled out completely. In the present context we only consider
 the effects of type I superconductivity on the Alfv\'en spectrum of 
torsional modes. The only other existing study of such a configuration (but with  
an additional toroidal component and assuming a discrete spectrum of modes) was
presented by \cite{Sotani2008b}.

We follow \cite{Aguilera2008} who give a description of different
axisymmetric magnetic field configurations. We neglect the effects of general
relativity on the magnetic field and the influence of the magnetic field on the
neutron star structure. This should be a valid approach, in particular, for
fields confined to the crust containing only a few per cent of the total 
stellar mass. Moreover, we assume that the magnetic field has relaxed to some 
equilibrium before the crust crystallized. After crystallization the magnetic field is
frozen into the crust. 
For simplicity,
we consider purely poloidal fields only. These models can be described by the type
III magnetic field of Section\,\ref{sec_bfield_general}. The
corresponding currents outside the crust determining the magnetic field
structure are the superconducting currents at the crust-core interface. 
Details on how to construct magnetic fields confined to the
crust are given in Appendix\,\ref{ap_crustfield}. 

We study different configurations, namely magnetic fields
matched to an exterior dipole, quadrupole, or octupole field. 
The amplitudes of the magnetic field at the surface of such configurations a
have different angular dependence. To compare the
results, we thus label the different configurations with their averaged magnetic
field strength at the surface of the star, and use this value as reference
magnetic field strength.

The field topology of the configurations and the corresponding spectra of the
purely Alfv\'en oscillations at $B=10^{14}\,$G are given in Fig.\,\ref{fig_crustfield}.
The parts of the Alfv\'en spectrum associated with a particular family of field
lines share the same colour in the graphs showing the spectrum and the
magnetic field configuration. If the magnetic field is confined to the crust
the maximum field strength is found near the crust-core interface. The maximum
strength for the dipole-like (quadrupole-like, octupole-like) configuration is
about $13$ ($8$, $8$) 
times the average value at the surface. This is a factor of a few stronger than the
crustal field of configurations penetrating the core.
Consequently, the structure of the magnetic field varies on smaller scales.
There are field lines which close inside the crust, e.g. near the
equator for the dipole-like (red lines in Fig.\,\ref{fig_crustfield}) and
octupole-like
(green lines) configurations, or at $\theta\sim60^\circ$ for quadrupole-like
and octupole-like configurations (see blue lines in
Fig.\,\ref{fig_crustfield}). 
Because of stronger magnetic fields and stronger field gradients on smaller
spatial scales, we need higher numerical grid resolution for our simulations (see below) 
and expect higher frequencies of the Alfv\'en oscillations.

In the following we will refer to magnetic field configurations matched
to an exterior dipole as configuration D, to an exterior quadrupole as Q, and to
an octupole as O, respectively.

\subsection{ Alfv\'en spectra with the semi-analytic model (without shear)}
\label{subsec_seman}

Let us first discuss the spectra obtained with the semi-analytic model of
\cite{Cerda2009} and \cite{Gabler2012} displayed in the lower panels of
Fig.\,\ref{fig_crustfield}. First, we note that for configurations matched to
 an exterior quadrupole and octupole, there are no turning points present inside
the continuum of the open field lines. Therefore, one would expect QPOs only at
the edges of the
different branches of the continua. For the dipole-like case we observe a
turning point only at the 
very high frequencies at field lines located around the closed field lines at
the equator (corresponding to the upper end of the red coloured part of the
spectrum). Second, the frequencies of the fundamental oscillation
of the short closed field lines ($>100Hz$) are higher than the frequencies we
observe for a magnetic field penetrating the core (a few Hz at
$10^{14}\,$G). The frequencies of the field lines near the polar axis are
similar to those obtained for a magnetic field
configuration penetrating the core. Note that there is a edge at $\theta=0$,
and that the corresponding frequencies are finite.

\subsection{Alfv\'en QPOs from simulations (without shear)}

The semi-analytic model shows that the spectra of magnetic fields confined to the crust 
are very steep near the edges and posses a turning point for configuration D
at very high
frequencies. Moreover, the field lines near the edges are packed very densely
along the crust-core interface 
leading to enhanced phase mixing. Therefore, we expect more difficulties to
observe QPOs
with significant amplitudes in our numerical simulations. As mentioned before, the
requirements on the grid resolution are more stringent, because of
the finer spatial structure of the magnetic field. 
The grid resolution employed for the time evolution of the three
 magnetic field configurations (at $B=10^{14}\,$G) is
$30\times 80$ ($r\times\theta$) for $[r_\mathrm{cc},
r_\mathrm{s}]\times[0,\pi]$. All models have been evolved for up to
$5\,$s.

\begin{figure}
\begin{center}
\begin{Large}
dipole-like D\\
\includegraphics[width=.4\textwidth]{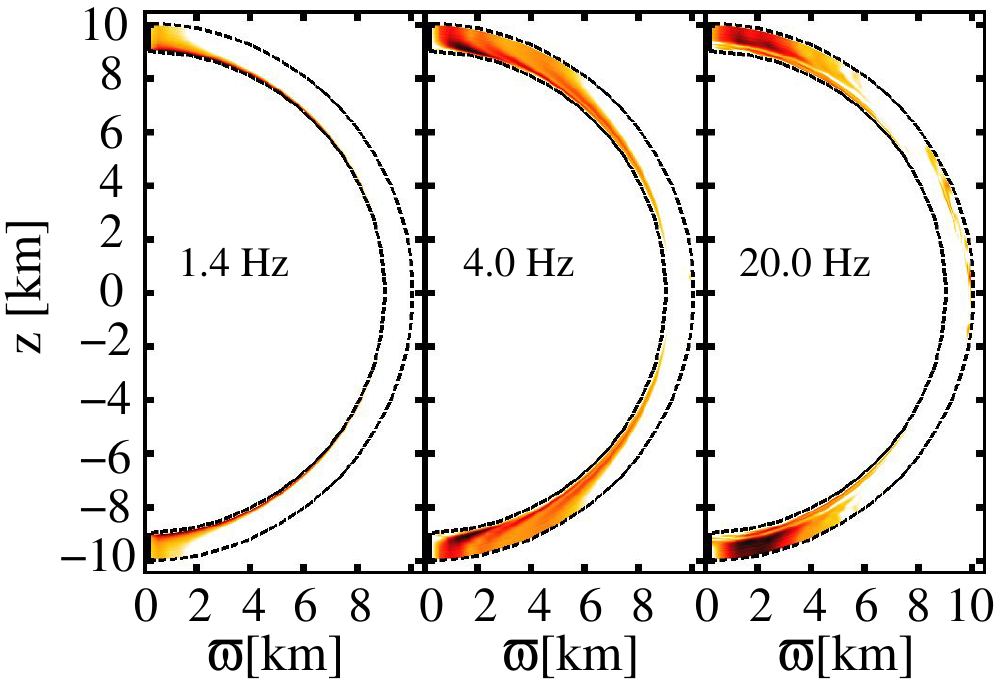} \\
quadrupole-like Q\\
\includegraphics[width=.4\textwidth]{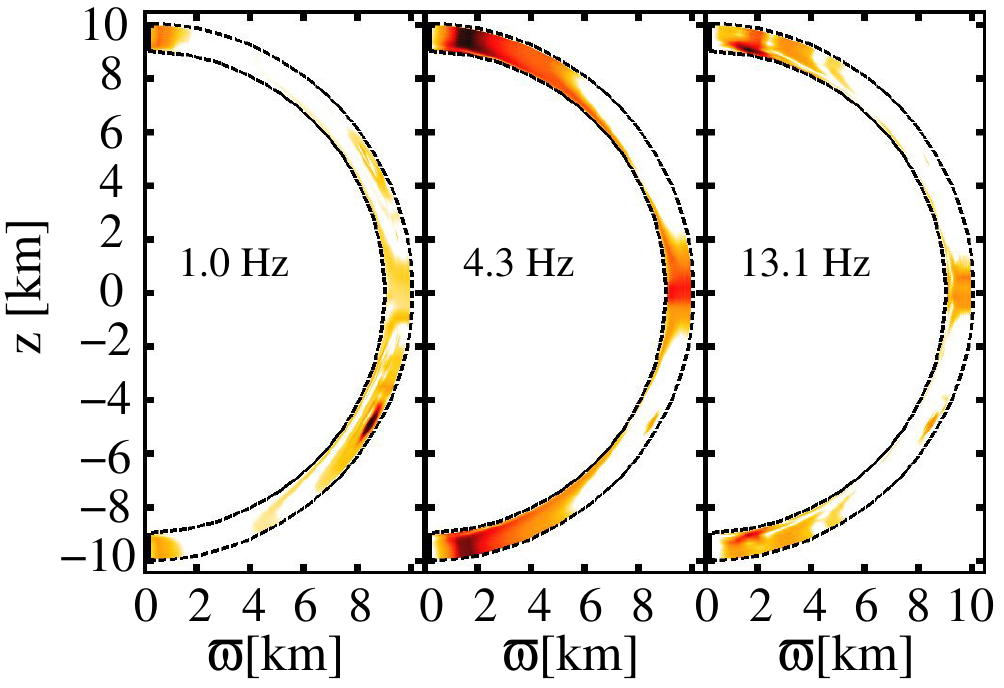} \\
octupole-like O\\
\includegraphics[width=.4\textwidth]{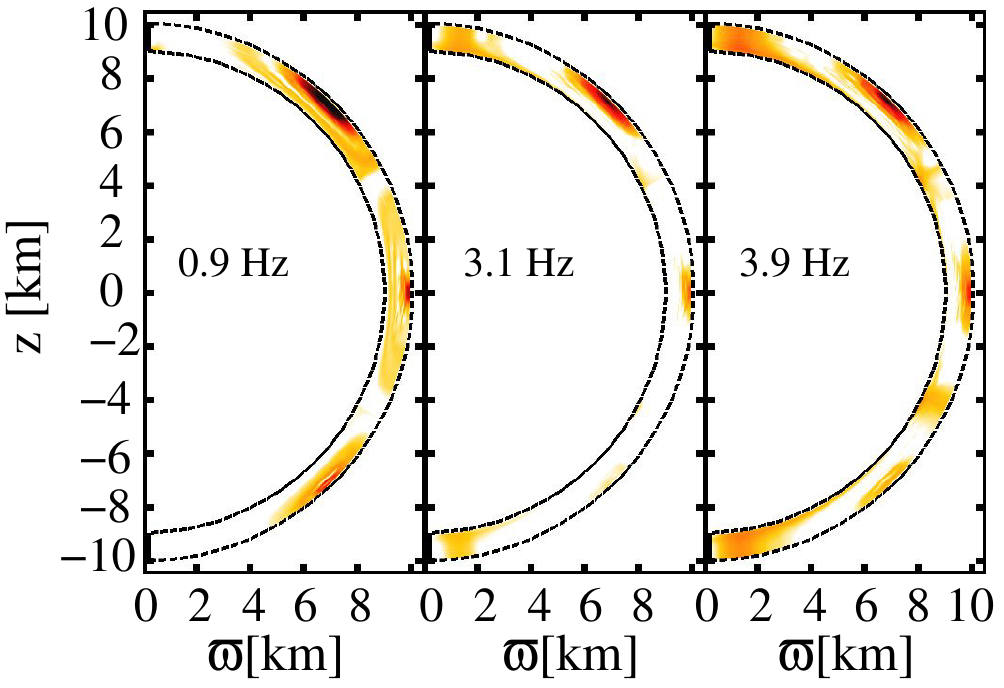}
\end{Large}
\end{center}
\caption{Spatial structure of the low-frequency QPOs obtained by a Fourier
analysis of different magnetic field
configurations. {\it Upper panels}: configuration 
D; {\it Middle panels:} configuration Q: {\it Lower
panels:} configuration O. The corresponding frequencies are given, too.
The colour scale ranges from white (minimum) via red to black (maximum).}
\label{fig_crustfield_QPOs}
\end{figure}

Indeed, the Fourier analysis displayed in Fig.\,\ref{fig_crustfield_QPOs} shows
less significant amplitudes of the QPOs than in the case of global magnetic
field configurations penetrating
the core. Nevertheless, we can identify some QPOs as local maxima of the Fourier amplitude. 
The QPOs for the dipole-like configuration (upper panels) have their maximum
amplitudes near the polar axis. This is expected, because the edge of the
continuum is located there. With the numerical resolution used we were not able
to see the other edge of the continuum at much higher frequencies, 
because the associated field lines near the equator 
have a very small spatial extension\footnote{
To resolve these smaller
scales, the grid resolution has to be refined significantly. However, we need to
evolve the models for several dynamical time scales of the low frequency
oscillations
around $1\,$Hz, which takes the order of several days of computing time
with the current resolution. Since we are mainly interested in the frequency
range around $30\,$Hz, we had to find a compromise between the computational
effort and the scanned frequency range.}.

The quadrupole-like magnetic field has closed field lines near $\theta\sim1.0$
rad (see 
Fig.\,\ref{fig_crustfield}). The corresponding frequencies of the continuum
are $>100\,$Hz. Unfortunately, the affordable grid resolution is too low to
analyze
potential QPOs for these field lines. Therefore, we find QPOs only for the
continuum
of the open field lines. These are best seen in the plot for $4.3\,$Hz in the
middle panel of Fig.\,\ref{fig_crustfield_QPOs}. Only open field
lines oscillate at the given frequency and have large Fourier
amplitudes near the polar axis and equator. Closed field lines, on the other
hand,
do not
participate in these oscillations, as the Fourier amplitude vanishes in the
corresponding regions.

When the magnetic field is matched to an exterior octupole, we find oscillations of the 
open field lines (best visible for $f=3.9\,$Hz in the bottom panel of Fig.\,\ref{fig_crustfield_QPOs}). The red-coloured field lines in the right panel of Fig.\,\ref{fig_crustfield} do not
extend to the equator, and the corresponding QPO has a
large amplitude around $\theta\gtrsim1.0$ rad. At 0.9 Hz we find oscillations
with significant amplitudes also for the field lines near the equator. 
At almost all frequencies there is a non-vanishing Fourier amplitude at the
closed field lines around $\theta\sim0.75$ and $\theta=\pi/2$, which is
probably caused by the too low grid resolution. 

In none of the cases considered is it possible to clearly associate the
observed QPOs with features of the semi-analytic spectrum obtained in Sec.
\ref{subsec_seman}, 
because the frequencies do not match perfectly. 
Nevertheless, a conservative interpretation would be that some QPOs which
are associated with
the open field lines near the polar axis correspond to the edges of the
continuum and
their overtones. The closed field lines
are expected to oscillate at frequencies above $100\,$Hz which we cannot confirm
because of lack of spatial resolution. Furthermore, we are mainly
interested in the frequency range below 100 Hz and models where the elastic
properties of the crust are included (see Section\,\ref{sec_crustfield_me}). 
Additionally, as a consequence of the finer spatial structure of the
magnetic field confined to the crust, there is a much stronger numerical
coupling of different field lines, which leads to enhanced damping and 
weaker QPOs.
We expect this weakness of our approach to be less important
in the coupled scenario, because for sufficiently low magnetic fields, the
oscillations will be dominated by the shear in the crust. The corresponding
modes are sufficiently resolved at the numerical resolution applied here.

\subsection{Coupled magneto-elastic oscillations}\label{sec_crustfield_me}

We now investigate the behaviour of coupled magneto-elastic oscillations of
the crust for the magnetic field configurations considered above.
Therefore, we performed a number of simulations with initial data
consisting of
a general perturbation with an angular dependence of the sum of the $l=2$ and
$3$ vector spherical harmonics at different magnetic field strengths.
In the following we will use the letter $l^\prime$ to label shear-mode-like
QPOs inside the crust, where $l^\prime$ is the number of nodes in
$\theta$-direction. For non-vanishing magnetic fields, these QPOs 
are no longer normal mode oscillations, but QPOs with time-varying
spatial structure. Therefore, they do not have the angular dependence of
individual vector spherical harmonics and 
they cannot be described by simple analytic functions.
The chosen grid resolution for the simulations depends on the selected model and on the radius 
at which the crust-core interface $r_\mathrm{cc}$ is located. The smallest resolution used
consisted of $60\times 80$ ($r\times\theta$) points for $[r_\mathrm{cc},
r_\mathrm{s}]\times[0,\pi]$. The integration time is $t\sim2\,$s at
$B=10^{14}\,$G. For stronger fields, the integration time scales
inversely with the magnetic field strength.

\begin{figure}
\includegraphics[width=.375\textwidth]{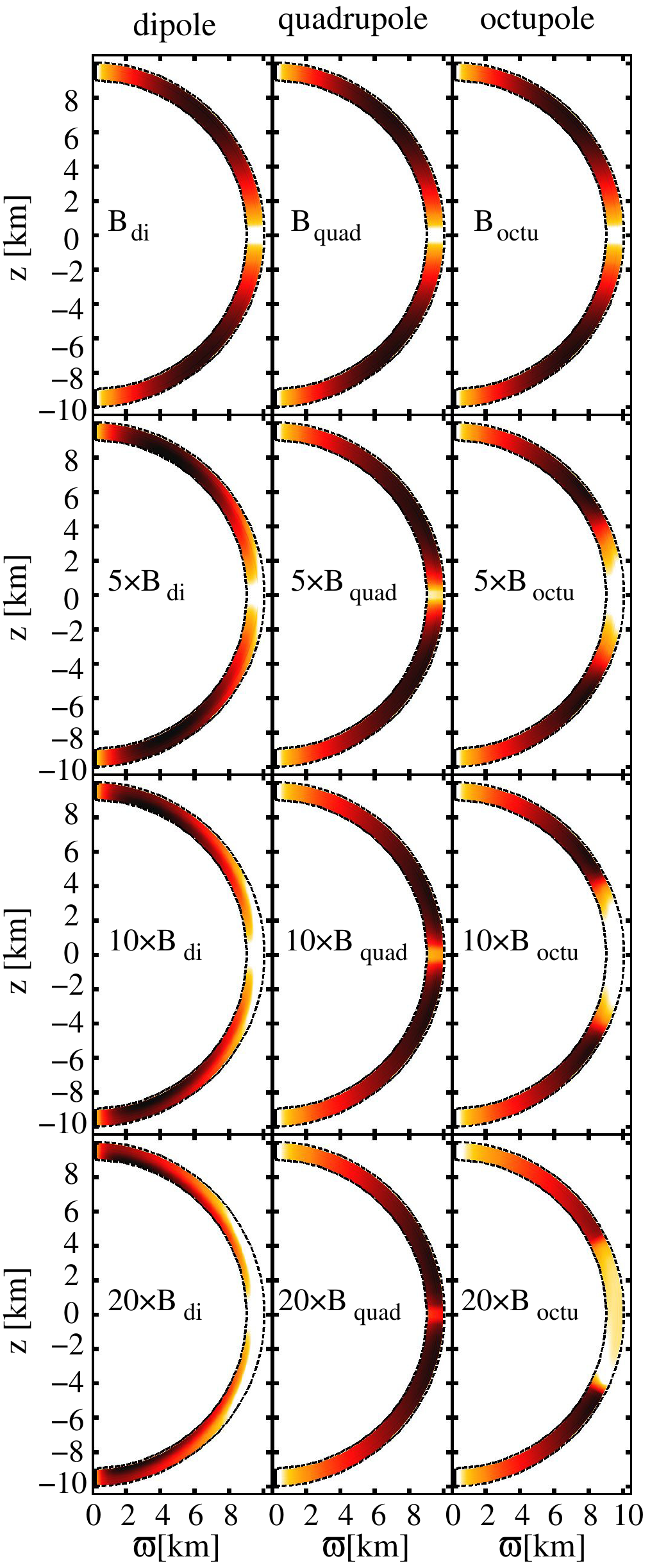} 
\caption{ Fourier amplitude of the $l^\prime=2$ crustal shear mode and its
magneto-elastic generalization for different magnetic field configurations and
different field strengths. The colour scale ranges from white (minimum) via red
to black (maximum).$B_\mathrm{di}=1.7\times10^{14}\,\mathrm{G}$,
$B_\mathrm{quad}=
1.9\times10^{14}\,\mathrm{G}$, and
$B_\mathrm{octu}=1.2\times10^{14}\,\mathrm{G}$, respectively.}
\label{fig_crustfield_modesI}
\end{figure}

The results of the Fourier analysis of these simulations are plotted in
Fig.\,\ref{fig_crustfield_modesI},  where we focus on the Fourier
amplitude of the magneto-elastic generalization of the $l^\prime=2$ shear mode
for different magnetic field strengths
and different field configurations (dipole-like, quadrupole-like, and
octupole-like).
The averaged surface magnetic field strengths are
$B_\mathrm{di}=1.7\times10^{14}\,\mathrm{G}$, $B_\mathrm{quad}=
1.9\times10^{14}\,\mathrm{G}$, and
$B_\mathrm{octu}=1.2\times10^{14}\,\mathrm{G}$, respectively.
When making general statements holding for all three magnetic field
configurations we use the label $B_\mathrm{x}$.

At $B=B_\mathrm{x}$ (first row of Fig.\,\ref{fig_crustfield_modesI}) the structure of 
the magneto-elastic QPOs is very similar to that of the purely crustal shear modes for all three
field configurations, and for all shear modes. The corresponding panels in
Fig.\,\ref{fig_crustfield_modesI} show how the purely crustal shear
modes approximately look like. For larger magnetic field strengths, the
assumed magnetic field configuration changes 
the structure of the shear modes. For example, in the presence of a
dipole-like surface 
field, the $l^\prime=2$ QPO is compressed towards those field lines entering the star closer 
to the pole. The stronger the magnetic field is the closer to the polar axis the
QPO are located (see at the leftmost panels of the figure from top to bottom).
This resembles the behaviour of the 
QPOs of purely Alfv\'en oscillations for this configuration (see
Fig.\,\ref{fig_crustfield_QPOs}). 
On the other hand, the QPO structure of quadrupole-like surface fields (central
panels) behaves in the opposite way, i.e. the maximum amplitude is
shifted towards the equator 
with increasing magnetic field strength. Finally, for an octupole-like
configuration (rightmost panels), 
the maximum Fourier amplitude remains at approximately its original position but it 
becomes narrower with increasing magnetic field 
strength. For the corresponding change in the frequency of crustal shear
modes with  $l^\prime>2$
(not shown in the figure) we observe a similar behaviour depending on the particular field 
configuration: The stronger the magnetic field, the more the structure of the QPO tends to 
resemble the structure of that field, and the shear modes are expelled from
regions of closed field lines (see e.g. the region close to the equator for
configurations D and O in Fig.\,\ref{fig_crustfield_QPOs}).

A common feature of all configurations is that with increasing
magnetic field strength the identification of QPOs becomes more difficult because
their amplitudes decrease. That is why some QPOs are not present at all field
strengths. In particular, we have not succeeded in identifying the following QPOs: $l^\prime=4$
at $B=10\times B_\mathrm{quad}$ and $B=20\times
B_\mathrm{quad}$, and $l^\prime=5$ at $B=20\times B_\mathrm{octu}$,
respectively. For even stronger
magnetic fields $B=50\times B_\mathbf{x}$ it is almost impossible to
identify the magneto-elastic generalizations of the crustal shear modes, as the
evolution is completely dominated by the magnetic field.

\begin{figure}
\begin{center}	
 \includegraphics[width=.47\textwidth]{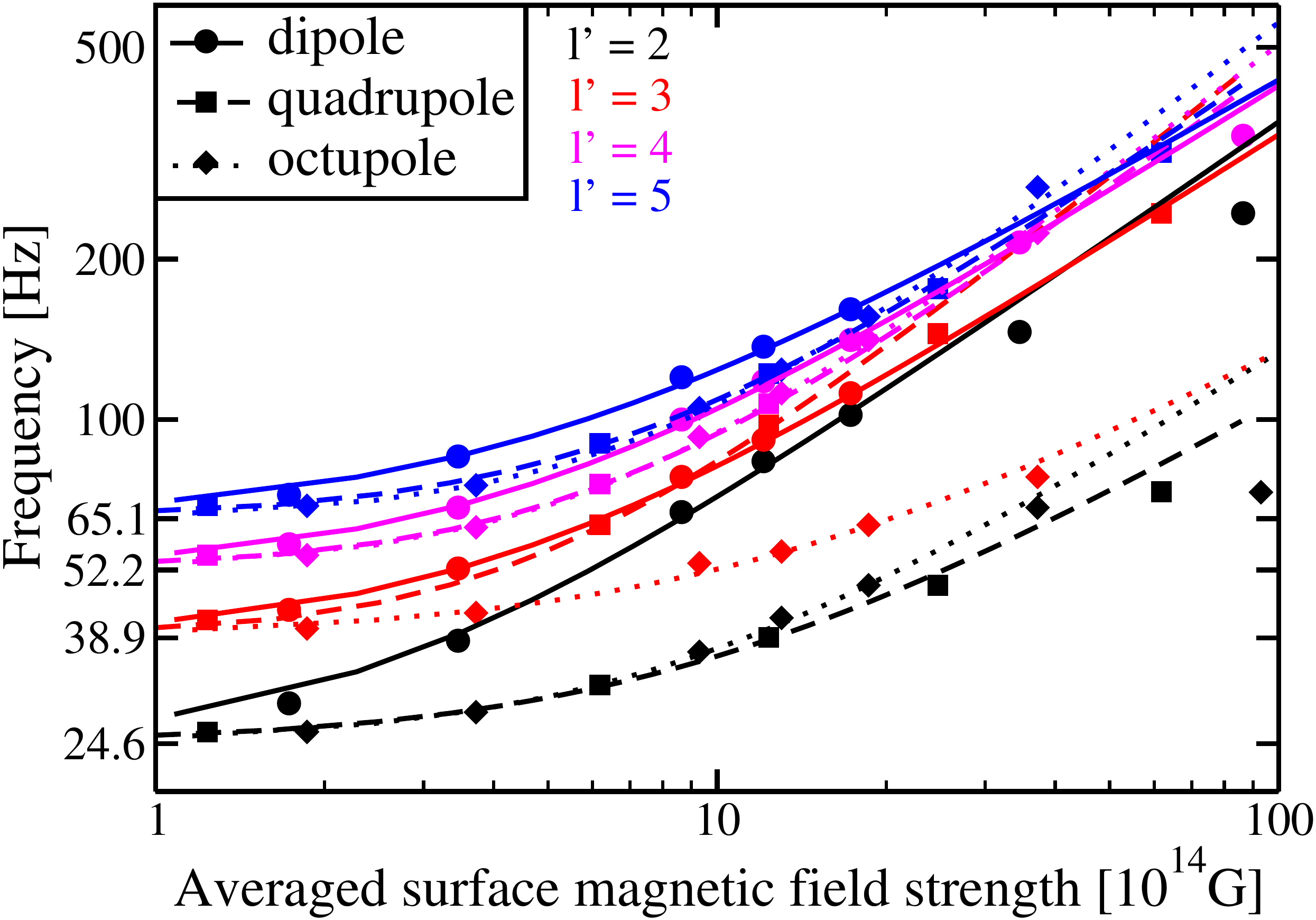}
\end{center}
\caption{Dependence of the frequencies of the $n=0$,
$l^\prime=\{2,~3,~4,~5\}$ QPOs on the magnetic field strength for different
magnetic field configurations which are confined to the crust. The purely
crustal shear mode
frequencies are given at the $y$-axis label as $24.6$, $38.9$, $52.2$, and
$65.1\,$Hz, respectively. The lines are fits according to
Eq.\,(\ref{eq_crustfield_fitII}), and the corresponding parameters are given in
Table\,\ref{tab_crustfield_fac}.}
\label{fig_crustfield_freqs}
\end{figure}

In order to understand how  the frequencies of the shear modes change,  when
they become QPOs in the
presence of the magnetic field we display in Fig.\,\ref{fig_crustfield_freqs} the QPO frequencies 
as a function of the averaged surface magnetic field strength. Different colours indicate different
QPOs associated with the zero-magnetic-field shear modes (black:
$l^\prime=2$;
red: $l^\prime=3$; magenta: $l^\prime=4$; blue: $l^\prime=5$). The
line style indicates the magnetic field configuration
(solid: dipole-like; dashed: quadrupole-like; dotted: octupole-like). At a
surface magnetic
field strength of about a few$\times10^{14}\,$G the frequencies of the QPOs
begin to deviate significantly from those of purely shear modes. For the
$l^\prime=\{2,3,4,5\}$ modes the zero-magnetic-field frequencies are $24.6$,
$38.9$, $52.2$, and $65.1\,$Hz, respectively, and are marked along the
$y$-axis

\cite{Sotani2008b} studied similar magnetic field
configurations and found an increase of the frequency of the QPOs with
increasing magnetic field. \cite{Sotani2007} give an approximate formula of the
dependence of the frequency on the magnetic field
\begin{eqnarray}\label{eq_crustfield_fitI}
 \frac{f_{l^\prime}(B)}{f_{l^\prime}^0} = \sqrt{1+a_{l^\prime}
\left(\frac{B}{B_\mu}\right)^2}\,,
\end{eqnarray}
where $f_{l^\prime}^0$ is the frequency of the purely shear eigenmode, and
$B_\mu=4\times10^{15}\,$G. When fitting the frequencies obtained in our
simulations with this formula we find a fairly inaccurate correspondence.
Therefore, we generalized Eq.\,(\ref{eq_crustfield_fitI}) to
\begin{eqnarray}\label{eq_crustfield_fitII}
 \frac{f_{l^\prime}(B)}{f_{l^\prime}^0} =
\sqrt{1+\left(\frac{B}{B_\mathrm{crit}}\right)^{b_{l^\prime}}}\,,
\end{eqnarray}
where $B_\mathrm{crit}$, is the magnetic field strength at which the influence
of the magnetic field becomes important. The parameters $B_\mathrm{crit}$ and
$b_{l^\prime}$ are given in
Table\,\ref{tab_crustfield_fac}, and the corresponding fits are
displayed in Fig.\,\ref{fig_crustfield_freqs}. Note that we only used the
values for $B\leq10\times B_\mathrm{x}$ for the fits because either we could
not find shear-mode-like oscillations for stronger fields or their identification was difficult.

The observed exponents deviate significantly from $b_{l^\prime}=2$, the majority
being $4/3\lesssim b_{l^\prime}\lesssim5/3$. For very strong magnetic
fields \citep[about several$\times10^{15}\,$G, see][]{Gabler2012} one would
expect a transition to Alfv\'en oscillations, where the frequency depends 
linearly on $B$, i.e. $b_{l^\prime}\rightarrow2$. However, in the
presence of such strong fields, shear oscillations no longer
exist. In the regime studied here we expect a significant influence of the shear
modulus such that the asymptotic regime of $b_{l^\prime}=2$ is not reached yet,
which may explain the deviation from $b_{l^\prime}=2$. 

\begin{table}
\begin{center}
\begin{tabular}{c | c | c c}
$l^\prime$ (nodes along $\theta$)& field & $B_\mathrm{crit}$ [$10^{14}\,$G]&
$b_{l^\prime}$\\
\hline
2&dipole-like&2.5&1.46\\
2&quadrupole-like&8.4&1.21\\
2&octupole-like&9.0&1.35\\\hline
3&dipole-like&4.1&1.35\\
3&quadrupole-like&4.7&1.70\\
3&octupole-like&13.1&1.14\\\hline
4&dipole-like&4.3&1.33\\
4&quadrupole-like&5.9&1.55\\
4&octupole-like&6.5&1.61\\\hline
5&dipole-like&4.6&1.22\\
5&quadrupole-like&6.7&1.46\\
5&octupole-like&7.8&1.63\\\hline
\end{tabular}
\caption{Parameters $B_\mathrm{crit}$ and $b_{l^\prime}$ of
Eq.\,(\ref{eq_crustfield_fitII}) for
different magnetic field configurations and different number of nodes
in $\theta$-direction $l^\prime$.
}\label{tab_crustfield_fac}
\end{center}
\end{table}

While there is no clear dependence of the exponent $b_{l^\prime}$ on the
magnetic field configurations, one should note that the critical surface field
$B_\mathrm{crit}$ is slightly lower for the dipole-like configuration 
than that for configurations Q and O for all QPOs and for all $l^\prime$.
This  can be
explained with the magnetic field structure. First, the configuration D has the
strongest magnetic field inside the crust ($\sim13\times B_\mathrm{surf}$) for
a given surface field compared to the other two ($\sim8\times B_\mathrm{surf}$).
Second, the $n=0$ shear modes\footnote{$n$ is the number of nodes in
radial direction.} travel predominantly in
$\theta$-direction. Therefore, the $\theta$-component of the magnetic field
should be the dominant one to enhance the mode propagation velocity and, hence, to 
increase the frequency of the QPOs.
In contrast to configuration Q and O, which both have more than one node of the
$\theta$-component of the magnetic field in $\theta$-direction, the
dipole-like configuration has only one node at the polar axis. Thus, the
averaged
absolute value of the $\theta$-component is stronger for configuration $D$ 
and we expect it to influence the magneto-elastic waves more strongly than in
case of the other two configurations.

We can extrapolate our results given by Eq.\,(\ref{eq_crustfield_fitII}) to different values 
of the shear modulus and test whether a change of this parameter can explain
the lowest observed frequencies of $18$, $26$, and $30\,$Hz in SGR 1806-20, as
proposed by \cite{Steiner2009}. This approach should be a very good
approximation for the $n=0$ modes, because their frequencies are mainly
determined by the shear modulus and depend only very weakly on additional
constraints from the EoS, such as the thickness of the crust
\citep{Samuelsson2007}.

If the shear modulus  $\mu_\mathrm{s}$ is divided by a factor $F$, the system
of equations (\ref{conservationlaw}), (\ref{eq_xi_dr}), and
(\ref{eq_xi_dtheta}) conserves its form if we rescale the
background magnetic field $B^k$ by the square root of this factor and
simultaneously multiply the time by $\sqrt{F}$:

\begin{eqnarray}
  \frac{\partial\sqrt{\gamma} S_\varphi }{\partial
\tilde{t}} &=& \frac{\partial \sqrt{-g} (b_\varphi \tilde{B}^k +
2\tilde{\mu}_\mathrm{s} \tilde{\Sigma}^k_{\,\,\varphi})}{\partial x^k} \,,\\
  \frac{\partial\sqrt{\gamma} {B}^\varphi }{\partial
\tilde{t}} &=& \frac{\partial \sqrt{-g} v^\varphi \tilde{B}^k}{\partial x^k}
\,,\\
{2}{g_{kk}g^{\varphi\varphi}}\frac{\partial\tilde{\Sigma}^k_{\,\,\varphi}}{
\partial\tilde { t} } &=&
\frac{\partial\tilde{\xi}^\varphi_{\,,k}}{\partial\tilde{t}} =
\frac{\partial v^\varphi
\alpha}{\partial x^k} \,,
\end{eqnarray}
where
$ \tilde{t} = {t} {\sqrt{F}}, \tilde{\Sigma}^k_{\,\,\varphi} =
{\Sigma}^k_{\,\,\varphi} {\sqrt{F}},
\tilde{B}^k = {B^k }/{\sqrt{F}}$, and
$\tilde{\mu}_\mathrm{s} = {\mu_\mathrm{s}}/{F}$, respectively.

The rescaling of the time coordinate implies an inverse rescaling of the
frequency.
Therefore, we obtain the expected result that in the absence of a magnetic field
the frequencies of the crustal shear modes scale as $1/\sqrt{F}$. A
decrease of the shear modulus by a factor $F$ leads to a corresponding decrease
of the QPO frequency by $\sqrt{F}$. 
Consequently, the critical magnetic field for an equilibrium model with a
rescaled shear modulus (divided by $F$) has to be divided by $\sqrt{F}$.
Knowing $B_\mathrm{crit}$ and $b_{l^\prime}$ for an equilibrium model we can
compute the frequency of the QPOs for different shear moduli by
\begin{equation}\label{eq_crustfield_fitIII}
 {f_{l^\prime}(B)} =
\frac{f_{l^\prime}^0}{\sqrt{F}} \sqrt{1+\left(\frac{B}{B_\mathrm{crit} /
\sqrt{F}}\right)^{b_{l^\prime} } } \, .
\end{equation}
The corresponding curves for the dipole-like background
configuration, the lowest QPO with $l^\prime=2$, and $F=\{2,4,8,16,100\}$ are
given in
Fig.\,\ref{fig_estimate_crustfield}. Clearly, the decrease of
the shear modulus leads to a decrease of $B_\mathrm{crit}$, i.e. the
magnetic field becomes dynamically dominate at lower magnetic field
strength than for large $\mu_\mathrm{s}$. With the dipole-like configuration we
are
not able to find oscillations at frequencies of $18\,$, $26$ or $30\,$Hz at
any value of the shear modulus up to $1/16$th of its tabulated value for
$B\gtrsim6\times10^{14}\,$G, which is the lower bound 
for the dipole-like magnetic field of the magnetars showing giant flares from
spin-down estimates. Only for the unrealistically small
value of $\tilde\mu_\mathrm{s}=\mu_\mathrm{s}/100$ we reach the observed
frequencies $f=30$ and $f=28\,$Hz for a magnetic field strength
of $B_{15}<1$.

\begin{figure}
\begin{center}	
\includegraphics[width=.475\textwidth]{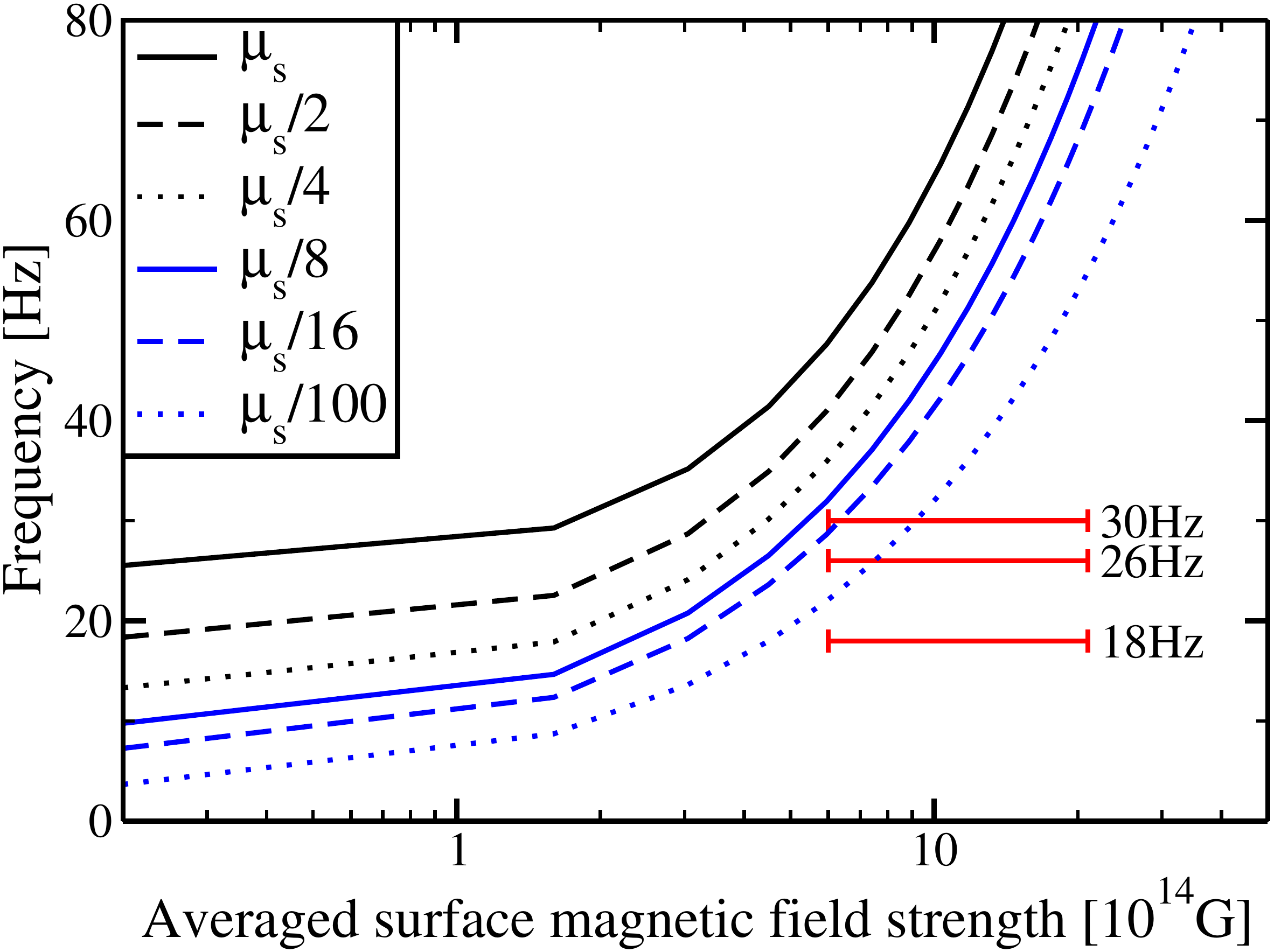}
\end{center}
\caption{Extrapolation of the QPO
frequencies using Eq.\,(\ref{eq_crustfield_fitIII}) for different values of the
shear modulus $\mu_\mathrm{s}$ in
the presence of a dipole-like magnetic field for the QPO with $l^\prime=2$. The
red
lines indicate the lowest observed frequencies of the QPOs for the range of
magnetic field strengths obtained using spin-down estimates
for the three magnetars showing giant flares 
($B_{15}=0.6\dots2.1$,
see http://www.physics.mcgill.ca/\texttildelow pulsar/magnetar/main.html).
}
\label{fig_estimate_crustfield}
\end{figure}

To summarize the findings of this section, we see a significant increase in the
frequency of the magneto-elastic generalization of the crustal shear modes with
increasing magnetic field
strength. The details of this increase, like the dependence on the magnetic
field strength or the spatial structure of the resulting QPOs, depend on the
particular magnetic field configuration. 
 For higher multipoles we find a weaker
increase of the QPO frequencies with the magnetic field strength than for 
dipole-like fields (see Fig.\,\ref{fig_crustfield_freqs}). However, the estimate
of the magnetic field strength with the spin-down formula gives a lower estimate
on the dipole-like component of the magnetic field. Therefore, the
results of this section can be seen as providing a lower estimate of the
frequency
increase if there are additional multipoles present in the magnetic field
structure.
At surface field strengths of $B_{15}\sim1$, expected from the
spin-down estimates, the frequencies of the crustal
shear modes are shifted to such high values that it is difficult to explain
the lowest frequencies of the QPOs observed in SGRs even for strongly reduced
shear moduli. 

This result shows that our simplified treatment of different effects of
the EoS on the magneto-elastic oscillations is valid in the particular case of
magnetic fields confined to the crust. The effect of the magnetic field is the
dominant one. Even when changing the shear modulus by a factor of about 100, we
barely reach the observed frequencies at $B\sim10^{15}\,$G. Any additional
correction on the structure of the crust or a more realistic calculation of the
shear modulus would have to change the shear mode frequencies by at least one
order of magnitude in the absence of the magnetic field in order to obtain
frequencies of $f\lesssim30\,$Hz in the presence of the field. Additionally, it
is not clear whether there would still exist shear-like QPOs in this case, i.e.
when the magnetic field is dominant\footnote{As we have
discussed previously in this section, the shear-like QPOs decrease in amplitude
with increasing magnetic field strength until they disappear completely.
Decreasing the shear modulus by a factor of 100, we would expect from
Fig.\,\ref{fig_crustfield_freqs} that the shear-like QPOs disappear at
$B\sim10^{15}\,$G.}.

\subsection{Different equilibrium models}
In addition to the APR+DH EoS used for all the results discussed so far in this paper, we
have  also investigated neutron star models of 1.4
solar masses obtained with the APR+NV, L+DH and L+NV EoS, respectively. For all
three EoS we
have performed simulations at magnetic field strengths of $B_{15}=0.1$,
$0.5$, $1.0$, $2.0$, and $5.0$, respectively. As discussed in the previous
section, it was highly problematic to identify
the QPOs corresponding to the pure shear modes in the zero field limit for the
strongest magnetic fields considered, because strong magnetic fields significantly
change the QPOs. In particular those QPOs having a maximum at the equator
($l^\prime=3$ and $l^\prime=5$) are difficult to detect for the dipole-like
field
configuration. The Alfv\'en oscillations along the closed field lines in this
region have very different structure and frequencies, and, thus, suppress
shear-like oscillations. 

\begin{table}
\begin{center}
\begin{tabular}{c | c | c c}
$l^\prime$ (nodes along $\theta$)& EoS & $B_\mathrm{crit}$ [$10^{14}\,$G]&
$b_{l^\prime}$\\
\hline
2&APRDH&2.5&1.35\\
2&APRNV&3.6&1.69\\
2&LDH&2.3&1.46\\
2&LNV&3.5&1.59\\\hline
3&APRDH&4.1&1.35\\
3&APRNV&5.3&1.49\\
3&LDH&3.4&1.36\\
3&LNV&5.2&1.48\\\hline
4&APRDH&4.3&1.33\\
4&APRNV&5.6&1.45\\
4&LDH&3.8&1.37\\
4&LNV&5.8&1.47\\\hline
5&APRDH&4.6&1.22\\
5&APRNV&6.0&1.41\\
5&LDH&4.1&1.27\\
5&LNV&6.0&1.39\\\hline
\end{tabular}
\caption{Parameters $B_\mathrm{crit}$ and $b_{l^\prime}$ of
Eq.\,(\ref{eq_crustfield_fitII}) for a magnetic field configuration matched
to an exterior dipole for different EoS and different number of nodes
in $\theta$-direction $l^\prime$, respectively.
}\label{tab_fac_eos}
\end{center}
\end{table}
The corresponding parameters $B_\mathrm{crit}$ and
$b_{l^\prime}$ for the fitting functions defined in Eq.
(\ref{eq_crustfield_fitII}) for $B_{15}\lesssim 1.5$
 are given in Table\,\ref{tab_fac_eos} for the set of EoS and magnetic field
amplitudes.
There is a clear, albeit weak, dependence of the two parameters on the QPO
structure characterized by $l^\prime$. For low $l^\prime$ the critical magnetic
field is weaker than for high $l^\prime$, and exactly the opposite holds for the
exponent $b_{l^\prime}$, which decreases with increasing $l^\prime$. The
influence of $B_\mathrm{crit}$ on the crustal EoS is of the same order of
magnitude as the difference between different $l^\prime$.
The DH EoS leads to approximately $30\%$ lower values of $B_\mathrm{crit}$ than
the NV EoS. This strong effect of the magnetic field for models with the DH EoS
can
be easily understood because the NV EoS leads to models with larger shear
moduli. The influence of the core EoS is less pronounced and one recovers very
similar $B_\mathrm{crit}$ and $b_{l^\prime}$ for the two examples studied here.

The dependence on the particular choice of the EoS is not very strong, and in
all cases we find that the frequencies increase significantly starting at a few
$10^{14}\,$G. Therefore, for these configurations and all EoS analyzed, we cannot
interpret the lowest observed QPO frequencies in terms of shear-like QPOs.
Additionally, without the identification of the fundamental shear mode, it is
not possible to explain the order of the higher frequency QPOs ($f>30$ Hz) with
shear modes. If the $l^\prime=2$ mode has a higher frequency of the order of the
frequencies of the higher $l^\prime$ modes does not match with observations.
Another possibility would be that the fundamental is not excited, but in this
case one would have to explain why this happens.

\section{Comparison with previous work}\label{sec_compare}

Several authors have pointed out the possible existence of long-lived 
discrete crustal modes in the gaps between the Alfv\'en continua of the core
(\cite{vanHoven2011}, \cite{Colaiuda2011}, and \cite{vanHoven2012}).
None of the simulations of \cite{Gabler2011letter}, \cite{Gabler2012}, and the 
present work show this behavior.
In \cite{Gabler2012} we performed a numerical experiment tailored
to find these crustal modes to no avail. In Section~\ref{sec_gaps} we
presented new numerical simulations to test whether pure crustal shear
oscillations
can survive in the gaps of the Alfv\'en continuum.
For this purpose we construct a particular model (F) with a flat spectrum
which has large gaps in between successive overtones. In none of our
simulations, however, we could find the crustal shear modes. Instead,
coupled magneto-elastic oscillations of the continuum were present.
In the following we compare these findings with those of previous work to
understand the reason for the apparent discrepancy.

\subsection{Comparison with Colaiuda \& Kokkotas}

\cite{Colaiuda2011} performed magneto-elastic simulations of axisymmetric torsional 
linear perturbations of magnetized neutron stars with a purely poloidal
background field.
They used tabulated EoS matching the core \citep[][among others]{Akmal1998}
with the crust \citep{Negele1973}. This combination is very similar to
 APRDH that is used in the present work, and the same combination has been
investigated in \cite{Gabler2012}. \cite{Colaiuda2011} used a constant
shear velocity in the crust that has a magnitude similar to the one used by us.
They did not consider any effect due to superfluidity. Therefore, their results
are directly comparable to those of \cite{Gabler2011letter}, \cite{Gabler2012},
and the present work (our work herafter).

The numerical simulations of \cite{Colaiuda2011} show that, even starting with
pure perturbations of the crust, {\it the energy of the oscillations quickly
flows from the crust towards the core exciting global oscillations}.  They
classify the oscillations as 'crustal modes', 'discrete Alfv\'en modes', 
and 'edges of the continuum'. The eigenfunctions of these oscillations
(Figs. 3, 4, 6, 7 and 8 in  \cite{Colaiuda2011}): (i) involve both crust and
core, (ii) are confined to a limited range of magnetic field lines, and (iii)
present nodes along the magnetic field lines but not across them. 
We also observe all three of these
features in the oscillation patterns of all of our magneto-elastic QPOs, 
which indeed do not significantly differ from the modes described 
in \cite{Colaiuda2011} despite of their different naming. 
Therefore, we think that these
oscillations correspond to excitations of the Alf\'en continuum.

\subsection{Comparison with van Hoven \& Levin}

\cite{vanHoven2012} performed simulations of the evolution of
axisymmetric torsional oscillations. Their background model is similar to 
our magnetar models. They used a tabulated EoS comparable to that of our
work, and the
same prescription for the shear modulus. The background magnetic field is purely poloidal
and corresponds to the configuration discussed in Section~\ref{sec_ring_current}.
However, the approach used by \cite{vanHoven2012} to solve
the magneto-elastic equations differs significantly from that in
\cite{Colaiuda2011} and in our work.
\cite{vanHoven2012} construct a model for magnetar oscillation
by coupling two sets of equations: one for the magneto-elastic oscillations of
the crust, and another one for the Alfv\'en oscillations of 
the core. In each region they search for an appropriate basis to expand
the evolution equations. The basis in the core corresponds to the
solution of the Sturm-Liouville problem for single magnetic field lines and zero
displacement at the core-crust interface. In the {\it coupled case} it is necessary to
introduce a new variable that is zero at the core-crust interface, because the
displacement does not have to vanish there, in general.
For the crust \cite{vanHoven2012} choose the basis to be the set of 
discrete modes of the magnetized crust with zero-traction condition (
$\xi^\varphi_{,r} = 0 $ ) at the base of the crust. However, when the fully
coupled system is considered, the displacement at the base of the crust
is neither limited to be zero nor to have a zero radial derivative. Therefore,
the fully consistent condition is not the zero traction condition, but the
continuous traction condition $\xi^\varphi_{\,\mathrm{core},r} = \left[ 1+
{\mu_\mathrm{S}}/{\Phi^4 (b^r)^2} \right]\xi^\varphi_{\,\mathrm{crust},r}$,
i.e. we find that the basis used in the crust by \cite{vanHoven2012} does 
not include all possible oscillations of the system,
but only the subset fulfilling $\xi^\varphi_{,r} = 0 $ at the core-crust
interface. This
effectively introduces a boundary condition at the core-crust interface which
artificially reflects part of the magneto-elastic waves.

Due to the restrictions of their boundary treatment 
\cite{vanHoven2012} neglect couplings between the crust and the core, in
particular those involving crustal modes
in the gaps of the continua, which do not show any interaction with the core in
their simulations. \cite{vanHoven2012} suggest that the
discrepancy between their and our works could be related
to the fact that they consider Alfv\'en waves coupling only to protons in the
superfluid core
while we do not. The effect of superfluidity on the magneto-elastic oscillations
is beyond the
scope of the present work and will be considered in future simulations to test
whether crustal modes can occur in that case. 

Actually, all QPOs of the
different works discussed in this section scale
more or less strongly with the magnetic field. Therefore, we prefer to call them
magneto-elastic QPOs rather than crustal modes, because the
latter expression gives the impression that the magnetic field does not
influence the properties of these QPOs and the crustal modes can be obtained as
in the field-free case, which is strictly not possible.

}

%
\section{Discussion}\label{sec_conclusions}

We studied the effect of different magnetic field 
configurations on the magneto-elastic oscillations of magnetars. For this
purpose we have constructed magnetized equilibrium configurations generalizing 
the {\small MAGSTAR} routine of the {\small LORENE}
library to include various descriptions of the current generating the magnetic
field: {\small MAGNETSTAR}\footnote{ Our generalization of {\small
MAGSTAR} is publicly available in the standard {\small
LORENE} library.}. The oscillation spectrum, dominated by the continuum of
the core, was studied by means of a semi-analytic model
\citep{Cerda2009,Gabler2012} and numerical GRMHD simulations. 

Table\,\ref{tab_spindown} summarizes the main results for the
magnetic field configurations
considered in this paper. For each model we identify the fundamental Alfv\'en QPO
as the turning point in the Alfv\'en continuum at the lowest frequency. We
identify
this mode with some of the observed QPOs, the $30\,$Hz oscillation of SGR
1806-20 and $28\,$Hz of SGR 1900+1, and we estimate the {\it equivalent dipole
magnetic field} strength of each
configuration to match both frequencies. We define the {\it equivalent
dipole magnetic field}, $\bar{B}$, as the magnetic field strength at the
surface of a Newtonian, uniformly magnetized sphere having the same dipole
magnetic moment\footnote{The value of $m$ is provided in the output of the
{\small LORENE} library, and the corresponding magnetic field strength can be
computed from $m=\bar{B} R^3$. For the ring current $m$ is given by
$m=IA$, where $I$ is the current and $A$ the area of the current loop.} $m$ as
the configuration we want to describe,
which can be directly compared with the magnetic field estimates from spin-down
measurements, through
$\bar B[\mathrm{G}]>3.2\times10^{19}\left(P[\mathrm{s}]\dot P\right)^{1/2}$
\citep{Lorimer2004}. 
Thus, the magnetic field estimates in Table\,\ref{tab_spindown} are directly
comparable to
estimates of the magnetic field from the spin down of the neutron star.
\begin{table}
\begin{center}
\begin{tabular}{c | c | c c}
Model&QPO&$\bar B_{15}^{28Hz}$&$\bar
B_{15}^{30Hz}$\\\hline
A0&U1&2.1&2.3\\
C$_{0.1}$&U1&1.3&1.4\\
C$_{10}$&U1&2.1&2.2\\
A1&U1&3.1&3.3\\
A1&U2&2.1&2.3\\
O&U1&3.6&3.8\\
O&U2&1.9&2.0\\
F&U1&2.4&2.5\\ \hline
R$_4$&U1&1.1&1.2\\
R$_4$&U2&0.85&0.93\\
R$_5$&U1&1.5&1.6\\
R$_5$&U2&0.80&0.85\\
R$_6$&U1&2.0&2.1\\
R$_7$&U1&2.7&2.8\\
R$_8$&U1&3.5&3.8\\\hline
$b_0=0$&U1&2.1&2.3\\
$b_0=1$&U1&2.2&2.4\\
$b_0=2$&U1&2.4&2.6\\
$b_0=5$&U1&2.9&3.1\\
$b_0=10$&U1&3.5&3.8\\
$b_0=20$&U1&3.7&4.0\\\hline
Q/D=0.1&U1&2.1&2.2\\
Q/D=0.5&U1&2.5&2.7\\
Q/D=0.5&U2&3.5&3.7\\
Q/D=0.5&U3&1.5&1.6\\
Q/D=1.0&U1$_N$&0.93&0.99\\
Q/D=1.0&U1$_S$&4.0&4.3\\
Q/D=2.0&U1$_N$&0.53&0.57\\
Q/D=2.0&U1$_S$&0.87&0.93\\
Q/D=5.0&U1$_N$&0.24&0.25\\
Q/D=5.0&U1$_S$&0.28&0.30\\
Q/D=10.0&U1$_N$&0.12&0.13\\
Q/D=10.0&U1$_S$&0.13&0.14\\
Q&U1&0.076&0.081\\\hline
\end{tabular}
\caption{Equivalent field strength $\bar B$ to match the frequencies at
$f=28$ and $f=30\,$Hz with the QPO indicated in the table. $\bar B$ is defined
by $m=\bar B R^3$ for a uniformly magnetized sphere, where $m$ is the magnetic
moment and $R$ is the radius of
the star. The QPOs $Un_X$ are the Upper QPOs of the open field
lines, $n$ indicates the number of the corresponding
turning points labelled from $n=1$ at the polar axis to maximum
$n=3$ at larger $\theta$, and $X$ gives the division into
northern ($X=N$) and southern ($X=S$) hemisphere, if there exist
different QPOs in both hemispheres.}\label{tab_spindown}
\end{center}
\end{table}

We summarize and discuss our main results next:

\begin{itemize}
 \item 
For poloidal, purely dipole-like configurations generated by a toroidal 
current inside the star (type II) we observe that the Alfv\'en continuum 
always contains 
turning points at the polar axis (U1). Some models (A1, O,
R$_4$, and R$_5$) have a second turning point (U2) in the region of open
field lines inside the star.
It seems very promising to identify some of the frequencies observed in the SGR
giant flares, which are in the integer relation of 1:3:5 to
each other\footnote{$30$, $92$, and $150\,$Hz for SGR 1806-20, and $28$, $84$,
and $155\,$Hz for SGR 1900+14}, with the fundamental Alfv\'en Upper QPO at
the polar axis and its second and fourth overtone. 
As can be seen in Table\,\ref{tab_spindown}, the required
equivalent field strength is in the range of $0.8$ to
$3.8\times10^{15}\,$G. This is consistent with the spin-down estimates of 
$\bar B_{15}>2.4$ for SGR 1806-20 and $\bar B_{15}>0.7$ for SGR 1900+14 
(where $\bar B_{15}$ is $\bar B$ in units of $10^{15}$ G), respectively.
 
\item 
The appearance of more than one turning point in the spectrum of some of the
configurations allows us to relate additional observed frequencies to the
torsional Alfv\'en oscillations. Model O was constructed such that we can match
all frequencies observed in SGR 1900+14. It is possible to identify the
$28$\,Hz with U1 at the polar axis, and the $84\,$Hz$=3\times28\,$Hz with its
second overtone. Additionally, the $53\,$Hz QPO could be associated to U2 away
from the polar axis at $\chi\sim5\,$km (see blue line in
Fig.\,\ref{fig_spectra}) and $155\,$Hz $\sim3\times53\,$Hz could be its second
overtone. Interpreting the $30\,$Hz QPO of SGR 1806-20 as U2, and the
$92\,$ and $150\,$Hz ones as the second and fourth overtones, one could
associate one of the lower frequencies, $18$ or $26\,$Hz, with U1.

\item 
We performed numerical simulations to test whether pure crustal shear
oscillations
can survive in the gaps of the Alfv\'en continuum, as it was proposed by
\cite{vanHoven2011}.
For this purpose we constructed a particular model (F) with a very flat spectrum
which has very
large gaps in between successive overtones. In none of our simulations we could
find crustal shear modes. Instead, 
 predominantly Alfv\'en oscillations of the continuum were present.

\item 
For mixed poloidal-toroidal magnetic field configurations (type I) where the
toroidal field is confined in the region of closed poloidal magnetic field lines 
inside the star, we estimate the part of the Alfv\'en spectrum of axial 
oscillations in the region of open field lines, {where the toroidal
component vanishes.} 
Qualitatively, the spectrum is similar to the one of purely poloidal fields
(type II) in that region.
With increasing field strength of the toroidal component, the poloidal field
strength near the closed field lines increases, too. This also means that the
surface field becomes stronger. Table\,\ref{tab_spindown} (3rd set) shows the
block for the different configurations $b_0$ we see that the necessary  {\it
equivalent dipole magnetic field}
strength to match the observed QPO frequencies in the spectrum increases with
increasing toroidal field, because of the stronger increase
of the
poloidal field strength close to the surface, while the field in the center is
only mildly affected.

\item 
Configurations with an additional quadrupole-like component have a richer
spectrum when this component is of comparable strength as the dipole-like one.
For $Q/D=0.5$, we find three turning points in the
spectrum. The additional turning points can produce oscillations which could
be used to match a larger number of observed frequencies. 
For a quadrupole dominated configuration $Q/D\gtrsim1.0$, we find
asymmetric oscillations in
both hemispheres of the neutron star. These provide a source for
additional QPO frequencies which could be matched to observations. Models with 
$Q/D>>1$, i.e. dominantly quadrupole-like configurations, are currently not
favoured,
since they require magnetic field strengths lower than standard spin-down
estimates.
The presence of a large quadrupole-like component in magnetars, comparable to
the poloidal component inside the star is not ruled out by our model. In fact 
the possible existence of this strong component could be related to
the amplification of
magnetic fields due to dynamo action, in initially rapidly rotating neutron
stars in the standard magnetar formation scenario
\citep{Duncan1992}. Once the dynamo stops, the magnetic field evolves mainly due to Hall drift
\citep{Goldreich1992} on a characteristic timescale
\begin{equation}
\tau_{\rm Hall} = 5\,10^4 \left( \frac{L}{1 {\rm km}}\right) \left( \frac{10^{16} {\rm G}}{B} \right) yr,
\end{equation}
 where $B$ is the typical magnetic field strength in the magnetar interior, and
$L$ is the typical length scale
of the magnetic field loops. Under such conditions, a quadrupole-like component
of similar strength as the dipole-like one could
still be present in magnetars, if it was present at birth.

\item
Finally, we investigated magnetic field configurations limited to the crustal
region. These configurations could be realized if the neutron star core is a
superconductor of type I. In this case the shear modes cannot be damped into
the core of the neutron star and they might survive sufficiently long to 
become observable.
However, our simulation results suggest that these models
provide no viable explanation for the observed frequencies in SGRs. We showed
that the influence of the magnetic field strength on the frequencies of the
magneto-elastic QPOs can be cast into a semi-empiric formula, which also
contains the effect of different shear moduli. With this relation, we are able to
scan a wide range of parameters of magnetic field strength and magnitude of the
shear modulus. The two possible types of QPOs are not favoured because:
i) On one hand, the structure and in particular the frequencies of
predominantly shear modes in the zero-magnetic-field case change
significantly with increasing magnetic field, such that the lowest observed
frequencies of
$f\lesssim30Hz$ cannot be reached for the currently expected values of the
shear modulus and
the magnetic field strength. ii) On the other hand, predominantly Alfv\'en
oscillations will be quickly damped by phase mixing due
to strong gradients in the spectra and an absence of turning points.
For different EoS we obtain similar results. As expected, the core EoS does not
play a large role in the determination of the critical magnetic field strength
at which the frequencies begin to increase significantly with the magnetic
field. The NV crust EoS leads to larger crusts with stronger shear moduli
compared to the DH EoS. This, in turn, causes the necessary magnetic field to
change the predominantly shear QPOs significantly to be up to
$2\times10^{14}\,$G stronger than for the DH EoS.
For quadrupole- or octupole-like dominated fields, the increase in frequencies
of predominantly shear QPOs is less strong compared to dipole-like fields.
However, from the neutron star spin-down measurement one has estimates of the
dipole-like component of $B_{15}\sim1$. At this dipole-like field strength our
previous arguments already apply and the frequencies are shifted to
$f\gtrsim30\,$Hz.
\end{itemize}

To summarize our findings, we have explored a large parameter space of possible
magnetic field configurations of neutron stars. We discovered additional
features, such as new turning points in the spectra for purely poloidal
configurations, which have the potential to explain more of the observed
QPO frequencies. Configurations with a toroidal component, which is
confined to the closed poloidal field lines, have very similar spectra compared
to those of purely poloidal fields. Furthermore, our results
do not favour the assumption that the magnetic fields are confined to the
crust.

Our magnetic field configurations do not include fields which have a toroidal
component extending throughout the whole neutron star, which may
be realized during a giant flare where strong toroidal fields are expected in
the exterior
\citep{Beloborodov2009, Fernandez2007, Nobili2008}. For such
configurations \cite{Colaiuda2011b} find discrete Alfv\'en oscillations.
However, these fields require currents at the surface of the neutron star,
which are difficult to model (and are set ad hoc in other studies). 
We postpone such considerations to future work.

The origin of the high frequency QPOs with $f=625\,$Hz or
higher remains open. In the magneto-elastic model the crustal $n=1$ shear modes
are damped less efficiently than the $n=0$ modes, however, they still disappear 
on timescales much too short to be observable \citep{Gabler2012, vanHoven2011}.
One potential alternative solution of this problem is related to
non-axisymmetric Alfv\'en oscillation of superfluid stars
\citep{Passamonti2012}.
Also oscillations of the magnetospheric field may play a role at these high
frequencies.

In a previous work \citep{Gabler2012} we have studied the effect of different
EoS for one particular magnetic field configuration. The differences in the
estimated magnetic field strength required to match observations that are caused
by changing the EoS are of the same order (factor of a few) as the differences
caused by assuming different magnetic field configurations. 
One of the main open questions in order to match observations
is the effect of superfluidity in the core of the neutron
star, which probably leads to lower (and thus probably more realistic) estimates
of the magnetic field strength \citep[see][for recent progress on
this topic]{Passamonti2012}. Furthermore, a model of how the 
coupling of the interior oscillations to the magnetosphere can lead
to a modulation of the X-ray emission needs to be developed.
We plan to address these two points in future studies. 

\section*{Acknowledgements}
We are grateful to S. Lander for fruitful discussions about the construction of
magnetic equilibria, and to Riccardo Ciolfi and Ioannis Contopoulos for useful
discussions. We are also grateful to Antonella Colaiuda and Kostas Kokkotas
for comments on the comparison of our results with their previous work. 
This work was supported by the Collaborative Research Center on Gravitational
Wave Astronomy of the Deutsche Forschungsgemeinschaft (DFG SFB/Transregio 7),
the Spanish {\it Ministerio de Educaci\'on y Ciencia} (AYA 2010-21097-C03-01)
the {\it Generalitat Valenciana} (PROMETEO-2009-103), the ERC Starting Grant CAMAP-259276, 
an IKY-DAAD exchange grant (IKYDA 2012)
and by CompStar, a Research Networking Programme of the European Science 
Foundation. N.S. also acknowledges support by an Excellence Grant for Basic
Research (Research Committee of the Aristotle University of Thessaloniki,
2012). 
The computations were performed at the {\it Servicio de Inform\'atica de la Universidad de Valencia}.

\appendix
\section{magnetic field configurations confined to the
crust}\label{ap_crustfield}
A more complete derivation of the following can be found in
\cite{Aguilera2008}. For axisymmetric fields it is possible to make the
following ansatz
\begin{equation}
 \mathbf{B}_\mathrm{poloidal} = {\boldnabla}\times(\mathbf{r}\times{\boldnabla}
\psi)\,,
\end{equation}
where $\psi=\psi(r,\theta)$ is a scalar whose angular part can be expanded
according to
\begin{equation}
 \psi(r,\theta) = C \sum_l \frac{P_l(\cos{\theta})}{r} S_l(r)\,,
\end{equation}
with $C$ being a normalization constant.
The poloidal magnetic field can thus be expressed as 
\begin{eqnarray}
 B_r &=& - \frac{B}{2 x^2} S_l(x) \frac{1}{\sin{\theta}}
\frac{\partial}{\partial\theta} \left(\sin{\theta}
\frac{\partial}{\partial\theta}P_l(\cos{\theta})\right)\,,\\
B_\theta &=& \frac{B}{2x} \frac{\partial}{\partial x}S_l(x)
\frac{\partial}{\partial\theta}P_l(\cos{\theta})\,,
\end{eqnarray}
where $x=r/r_\mathrm{s}$ and $B$ is a constant related to $C$. For dipolar
fields $C=r_\mathrm{s}^2 B / 2$. The $\varphi$-component of the force-free
condition
${\boldnabla}\times\mathbf{B}= \mu \mathbf{B}$ leads to a Riccati-Bessel
equation for
$S_l(x)$ \citep[see][]{Aguilera2008}. $\mu$ is a parameter related to the
currents
maintaining the magnetic
field. For different $l$, we have 
\begin{equation}\label{eq_pot_crustfield}
S_l(x) =  a_l \mu r_\mathrm{s} x\,j_l(\mu r_\mathrm{s} x) + b_l \mu r_\mathrm{s}
x\,n_l(\mu r_\mathrm{s} x)\,,
\end{equation}
where $a_l$ and $b_l$ are parameters, and $j_l$ and $n_l$ are the spherical
Bessel
functions of first and second kind. The parameters $a_l$ and $b_l$ can be
obtained by matching to an exterior solution of the magnetic field. 

For a given $l$ both have to be determined by the boundary
conditions for $B_r$ and $B_\theta$ at the surface. For a dipole external field
we get 
\begin{eqnarray}
 a_1&=&\cos{\mu r_\mathrm{s}}\,,\\
b_1&=&\sin{\mu r_\mathrm{s}}\,,
\end{eqnarray}
 while for the quadrupole external field 
\begin{eqnarray}
 a_2&=&3 \frac{ \sin{\mu r_\mathrm{s}}}{(\mu r_\mathrm{s})^2} - \sin{\mu
r_\mathrm{s}} - 2 \frac{\cos{\mu r_\mathrm{s}}}{(\mu r_\mathrm{s})} +
3\frac{\cos{\mu r_\mathrm{s}}}{(\mu r_\mathrm{s})^3}\,,\\
b_2&=&-3 \frac{\cos{\mu r_\mathrm{s}}}{(\mu r_\mathrm{s})^2} + \cos{\mu
r_\mathrm{s}} - 2 \frac{\sin{\mu r_\mathrm{s}}}{(\mu r_\mathrm{s})} +
3\frac{\sin{\mu r_\mathrm{s}}}{(\mu r_\mathrm{s})^3}\,.
\end{eqnarray}
The octupole field can be matched by
\begin{eqnarray}
 a_3&=&\cos{\mu r_\mathrm{s}} - 5 \frac{\sin{\mu r_\mathrm{s}}}{\mu
r_\mathrm{s}}
-15 \frac{\cos{\mu r_\mathrm{s}}}{(\mu r_\mathrm{s})^2} + 30 \frac{\sin{\mu
r_\mathrm{s}}}{(\mu
r_\mathrm{s})^3} \nonumber\\
&&+30\frac{ \cos{\mu r_\mathrm{s}}}{(\mu
r_\mathrm{s})^4}\,,\\
b_3&=&\sin{\mu r_\mathrm{s}} + 5 \frac{\cos{\mu r_\mathrm{s}}}{(\mu
r_\mathrm{s})} - 15\frac{\sin{\mu r_\mathrm{s}}}{(\mu
r_\mathrm{s})^2} - 30 \frac{\cos{\mu r_\mathrm{s}}}{(\mu r_\mathrm{s})^3}
\nonumber\\
&&+30 \frac{\sin{\mu r_\mathrm{s}}}{(\mu r_\mathrm{s})^4}\,.
\end{eqnarray}

It remains to determine the value of $\mu$. For magnetic fields confined to the
crust $B_r=0$ at the crust-core interface. This translates
into the condition $S_l|_{r_\mathrm{cc}} = 0$, and hence for $l=1$ into
\begin{equation}
 \tan{[\mu(r_\mathrm{cc} - r_\mathrm{s})]} - \mu r_\mathrm{cc} =0\,,
\end{equation}
which has to be solved numerically. For the quadrupole- and octupole-like
configurations the corresponding equations are more complicated and are not
given here.

The particular form of $A(x)$ in general does not have to coincide with
the solution in terms of spherical Bessel functions presented
above. But because of arbitrariness it is advantageous to use an analytical
description here.
%

\bibliographystyle{mn2e}
\bibliography{magnetar.bib}

\end{document}